\documentclass[aps,prd,10pt,onecolumn,nofootinbib,showpacs,showkeys,notitlepage]{revtex4-1}
\usepackage{latexsym,graphicx,amssymb,amsmath,mathrsfs}
\usepackage{setspace,bm}              
  
\newcommand{\be}{\begin{equation}}
\newcommand{\ee}{\end{equation}}
\newcommand{\bea}{\begin{eqnarray}}
\newcommand{\eea}{\end{eqnarray}}
\newcommand{\bse}{\begin{subequations}}
\newcommand{\ese}{\end{subequations}}

\newcommand{\tr}{\textrm{tr}}
\newcommand{\tn}[1]{{\textnormal{#1}}}
\newcommand{\grdstate}[1]{\left\langle #1 \right\rangle }
\def\p{{\bf p}}
\def\k{{\bf k}}
\def\q{{\bf q}}
\newcommand{\slk}{\mbox{\,\slash \hspace{-0.5em}$k$}}

\newcommand{\rt}[1]{{}}

\newcommand{\un}{\mbox{$\mathbf{1}$ \hspace{-0.91em}
    \raisebox{0.05em}[0pt]{$\shortmid$} \hspace{-0.895em} \raisebox{0.235em}[0pt]{$\shortmid$}}}

\begin{document}
\allowdisplaybreaks{

\title{Influence of the Polyakov loop on the chiral phase transition 
in the two flavor chiral quark model}

\author{G. Mark{\'o}}
\email{smarkovics@hotmail.com}
\affiliation{Department of Atomic Physics, E{\"o}tv{\"o}s University,
H-1117 Budapest, Hungary}

\author{Zs. Sz{\'e}p}
\email{szepzs@achilles.elte.hu}
\affiliation{Statistical and Biological Physics Research Group
of the Hungarian Academy of Sciences, H-1117 Budapest, Hungary}   

\begin{abstract}
  The $SU(2)_L\times SU(2)_R$ chiral quark model consisting of the
  $(\sigma,\vec\pi)$ meson multiplet and the constituent quarks
  propagating on the homogeneous background of a temporal gauge field
  is solved at finite temperature and quark baryon chemical potential
  $\mu_q$ using an expansion in the number of flavors $N_f$, both in
  the chiral limit and for the physical value of the pion mass.
  Keeping the fermion propagator at its tree-level, several
  approximations to the pion propagator are investigated. These
  approximations correspond to different partial resummations of the
  perturbative series.  Comparing their solution with a
  diagrammatically formulated resummation relying on a strict
  large-$N_f$ expansion of the perturbative series one concludes that
  only when the local part of the approximated pion propagator resums
  infinitely many orders in $1/N_f$ of fermionic contributions a
  sufficiently rapid crossover transition at $\mu_q=0$ is achieved
  allowing for the existence of a tricritical point or a critical
  end point in the $\mu_q-T$ phase diagram. The renormalization
  and the possibility of determining the counterterms in the
  resummation provided by a strict large-$N_f$ expansion are
  investigated.
\end{abstract}


\maketitle

\section{Introduction}

The low-energy effective models of the QCD, such as the
Nambu--Jona-Lasinio (NJL) model \cite{nambu61} and the linear sigma
model (also called chiral quark or quark-meson (QM) model)
\cite{gell-mann60} are based on the global chiral symmetry of the
QCD. They proved to be very useful in qualitative understanding of
many aspects related to the spontaneous breaking of the chiral
symmetry and its restoration at finite temperature and density, but
share as a major drawback the lack of the confinement property. As a
consequence of the absence of gluonic effective degrees of freedom and
due to the lack of color clustering \cite{megias06} there are
unsuppressed contributions of constituent quarks in the
low-temperature phase. Both features, which are in fact related, alter
the reliability of the quantitative thermodynamic predictions of these
models, such as the equation of state or the location of the critical
end point (CEP) in the $\mu_q-T$ phase diagram.

Since the QCD phase transition involves both the restoration of chiral
symmetry measured by the evaporation of the chiral condensate
$\grdstate{\bar\psi\psi}$ and the liberation of quarks and gluons
encoded in the change of the Polyakov loop $\Phi$, much effort was
devoted to understand the relation between the chiral and
deconfinement phase transitions. An argument by Casher \cite{casher79}
states that in the vacuum, that is at $\mu_q=T=0,$ confinement implies
the breaking of the chiral symmetry. The connection between
$\grdstate{\bar\psi\psi}$ and $\Phi$ is revealed by the spectral
density of the Dirac operator $\rho(\lambda)$.  In the vacuum the
Banks-Casher relation $\rho(0)=\grdstate{\bar\psi\psi}/\pi$
\cite{banks80} states that the spectral density of the Dirac operator
in the deep infrared is proportional to the quark condensate. Finite
temperature lattice studies at $\mu_q=0$ show that the connection
between chiral symmetry and confinement suggested by Casher's
argument holds. The infrared part of $\rho(\lambda)$ undergoes a
pronounced change as one crosses from the confined to the deconfined
phase \cite{edwards00,kovacs08}.  As shown recently in
\cite{gattringer06}, the phase of the Polyakov loop $\Phi,$ which can
be expressed as a spectral sum of eigenvalues and eigenvectors of the
Dirac operator with different boundary conditions, receives its main
contribution from the infrared end of $\rho(\lambda).$ It is generally
true for both $N_c=2$ and $N_c=3$ that at high temperature the fermion
determinant favors the sector where the Polyakov loop lies along the
positive real axis \cite{edwards00,kovacs08}. For this type of
configuration in which the phase of $\Phi$ vanishes the chiral
symmetry is restored, because a sizable gap develops in the spectral
density of the Dirac operator which implies in view of the
Banks-Casher relation $\grdstate{\bar\psi\psi}=0.$ For configurations
in which the phase of the Polyakov loop is not vanishing, the chiral
symmetry is not restored, as observed in the lattice study of the quenched
QCD \cite{chandrasekharan96a} and also by using a random matrix model
calculation \cite{stephanov96}.

Casher's argument suggests that the temperature $T_d$ for the
deconfinement phase transition is somewhat lower than the restoration
temperature $T_\chi$ of the chiral symmetry. As explained in
\cite{glozman09} at finite density Casher's argument could fail, so that it
does not contradict the existence of a dense phase in which at a given
temperature chiral symmetry is restored while quarks remains confined.
Such a phase can exist inside the so-called quarkyonic phase, which
was suggested as a new phase of the QCD at finite temperature and
density, based on its existence within a large-$N_c$
analysis~\cite{mcLerran07}.

In the Polyakov-loop extended NJL model (PNJL) where the coupling of
the Polyakov loop to the quark sector is achieved by the propagation
of the quarks on a constant temporal gauge field background, the
simultaneous crossover-type transition of deconfinement and chiral
restoration was obtained \cite{fukushima04b}.  As shown in
\cite{meisinger96,chandrasekharan96b} this model is able to reproduce
the main features of the quenched lattice result of
\cite{chandrasekharan96a}. The phase transition was recently
intensively investigated in the PNJL model with two
\cite{hansen07,abuki08,kashiwa08,costa09a,sakai09} and three flavors
\cite{ciminale08,fu2008,costa09b}, also in the nonlocal formulation
of the model \cite{sasaki07,dumm10}.  The interplay between chiral and
deconfinement transitions was investigated in the PNJL model using
large-$N_c$ expansion in \cite{mcLerran09}.

By coupling the Polyakov loop to the quark degrees of freedom of the
QM model the thermodynamics of the resulting Polyakov quark-meson
model (PQM) was studied for two
\cite{schaefer07,tuominen08,nakano10,skokov10,skokov10b,herbst10} and three
quark flavors \cite{schaefer10,mao10,gupta10}. The effect of a strong
magnetic field expected to be generated in the LHC in noncentral
high-energy heavy ion collisions on the chiral and deconfining phase
transitions was studied recently in \cite{mizher10} within the PQM
model with two flavors.  The possibility of coupling the Polyakov loop
to meson models without quarks was considered in
\cite{sannino04,fraga}.

The coupling of the Polyakov loop to the chiral effective models
mimics the effect of confinement by statistically suppressing at low
temperature the contribution of one- and two-quark states relative to
the three-quark states. This feature makes the Polyakov-loop extended
effective models more appropriate for the description of the
low-temperature phase and for quantitative comparison with the
thermodynamic observables on the lattice
\cite{weise08,schaefer07,schaefer10}.  Better agreement is expected up
to $T\simeq (1.5-2) T_c$ above which the transverse gluonic degrees of
freedom dominate in thermodynamic quantities, such as the pressure,
over the longitudinal ones represented by the Polyakov loop.

Despite this success, one should keep in mind that the solution of the
Polyakov-loop extended effective models is mainly obtained in the
lowest one-loop (trace-log) order of the fermionic sector, hence
studying their stability against inclusion of higher loops would be
certainly of interest. Some approximations to the PQM model
\cite{schaefer07,tuominen08,schaefer10,mao10,gupta10} neglect the
fermionic vacuum fluctuations and by treating the mesonic potential at
tree level, completely disregard quantum effects in the mesonic
sector. The effect of including the quantum fluctuation in the PQM
model was recently studied in \cite{nakano10,skokov10,herbst10} 
using functional renormalization group methods.

In this work we would like to address two questions using large-$N_f$
approximation to the $SU(2)_L\times SU(2)_R\simeq O(4)$ model.  The
first one is to what extent the inclusion of the Polyakov loop
modifies the $\mu_q-T$ phase diagram obtained previously in
\cite{toni04} in the chiral limit of the two flavor QM using the
large-$N_f$ approximation. The second one concerns the effect of
different partial resummations on the quantitative results. To this
end several approximate resummations of the perturbative series will
be investigated and the obtained results compared.

The paper is organized as follows. In Sec.~II we overview some basic
facts about the Polyakov loop, including different forms of the
effective potential and we introduce and parametrize the PQM model,
presenting also the approximations exploited for its solution. The
renormalization of the model and the determination of the counterterms
is discussed in Sec.~III. In Sec.~IV we present the numerical results
on the $\mu_q-T$ phase diagram obtained in the chiral limit and for
the physical value of the pion mass by using different forms of the
Polyakov-loop effective potential and various approximations to the
resummed pion propagator. Section V is devoted to discussion and
summary.

\section{The PQM model within a large-$N_f$ approximation 
\label{sec:model}}

\subsection{The Polyakov loop as an order parameter\label{ss:POP}}
We shortly review a few well known facts about the Polyakov loop
incorporated as a new effective degree of freedom in the
chiral quark model. This is usually done by considering the
propagation of quarks on the homogeneous background of a temporal
gauge field $A_0(x).$ At finite temperature $T=1/\beta$, after 
analytical continuation to imaginary time $t\to i\tau,$ $A_0\to i A_4,$ 
the temporal component of the Euclidean gauge field
$A_4$ enters in the definition of the Polyakov-loop operator (path
ordered Wilson line in temporal direction) $L(\vec x)$ and its
Hermitian (charge) conjugate $L^\dagger(\vec x)$
\be
L(\vec x)={\cal P} \exp\left[i\int_0^\beta d\tau A_4(\tau,\vec x) \right],
\qquad
L^\dagger(\vec x)={\cal P} \exp\left[-i\int_0^\beta d\tau A_4^*(\tau,\vec x) 
\right],
\label{Eq:Polyakov_op}
\ee
which are matrices in the fundamental representation of the $SU(N_c)$ 
color gauge group ($N_c=3$). In the so-called Polyakov gauge, the temporal 
component of the gauge field is time independent and can be gauge rotated 
to a diagonal form in the color space 
$A_{4,d}(\vec x)=\phi_3(\vec x) \lambda_3+\phi_8(\vec x) \lambda_8$ 
\cite{reinhardt97,ford98,schaden05},
where $\lambda_3,\lambda_8$ are the two diagonal Gell-Mann matrices.
Then the Polyakov-loop operator simplifies
\bea
L(\vec x)&=&
\tn{diag}(e^{i\beta\phi_+(\vec x)},
e^{i\beta\phi_-(\vec x)},e^{-i\beta(\phi_+(\vec x)+\phi_-(\vec x))}),
\label{Eq:Polyakov_op_diag}
\eea
where $\phi_\pm(\vec x)=\pm \phi_3(\vec x)+\phi_8(\vec x)/\sqrt{3},$
with a similar form for the conjugate $L^\dagger(\vec x).$
 
Topologically nontrivial gauge transformations 
$U(\tau,\vec x)\in SU(N_c)$ that are periodic up to a twist, that is 
$U(\tau+\beta,\vec x)=z U(\tau,\vec x),$ were introduced in \cite{thooft}, 
where $z$ is an element of the center of the $SU(N_c)$ group which is 
isomorphic with
$\mathbb{Z}_{N_c}=\{z|z=\exp(2\pi n i/N_c), n=0,1, \dots, N_c-1\},$
the cyclic group of order $N_c.$
Under such transformations the color trace of the Polyakov-loop 
operator and its conjugate, that is $l(x)=\tr_c L(\vec x)/N_c$ and 
$l^\dagger(x)=\tr_c L^\dagger(\vec x)/N_c,$ are transformed by an element of 
the center: $l(x)\to z l(x), l^\dagger(x)\to z^* l^\dagger(x).$ 
Consequently, in the pure gauge theory, which has an exact 
$\mathbb{Z}_{N_c}$ global symmetry, the thermal expectation value 
of the traced Polyakov-loop operator and its conjugate,
\bea
\Phi(\vec x)=\frac{1}{N_c} \grdstate{\tr_c L(\vec x)},\qquad 
\bar\Phi(\vec x)=\frac{1}{N_c} \grdstate{\tr_c L^\dagger(\vec x)},
\label{Eq:Polyakov_field}
\eea
are order parameters for the center symmetry and must vanish if the 
symmetry is unbroken. However, the Polyakov 
loop $\Phi(\vec x)$ and its conjugate $\bar\Phi(\vec x)$ can acquire 
a nonvanishing value, signaling the spontaneous breaking of 
the $\mathbb{Z}_{N_c}$ symmetry. These complex quantities can be regarded as
order parameters of the deconfinement phase transition, because the
free energy of a heavy (static) quark-antiquark pair with spatial separation
$\vec r=\vec x-\vec y$ is related to the expectation value of the
correlator of the traced Polyakov-loop operator for which cluster
decomposition is expected to hold at infinite separation
\be
\exp\left[-\beta F_{q\bar q}(\vec r)\right]=
\frac{1}{N_c^2} \grdstate{L(\vec x) L^\dagger(\vec x)} \to
\Phi(\vec x)\bar\Phi(\vec y).
\label{Eq:Polyakov_cluster}
\ee
When $\Phi(\vec x),\bar\Phi(\vec y)=0$ then 
$F_{q\bar q}(\vec r)\to \infty,$ and when $\Phi(\vec x),\bar\Phi(\vec y) \ne 0$ 
then $F_{q\bar q}(\vec r)$ is finite, which means confinement and 
deconfinement, respectively \cite{mclerran81,svetitsky86}.

In the presence of dynamical fermions the $\mathbb{Z}_{N_c}$ symmetry
is not exact anymore. Nevertheless, the Polyakov loop gives through its
distribution information about the confinement (low $T$) or
deconfinement phase (high $T$) of the system both in a canonical or
grand-canonical formulation of the QCD \cite{faber95,kratochvila06}.
Since its absolute value can be related to the free energy difference
between two systems, one containing the quark-antiquark source pair and the
other not containing it, by renormalizing the free energy a
renormalized Polyakov loop can be defined \cite{peter02}, which
provides information on the temperature of the deconfinement phase
transition.

\subsection{The mean-field Polyakov-loop potentials \label{ss:PEP}}

In the mean-field approximation $\Phi(x)$ and $\bar\Phi(x)$ are
replaced by $x$-independent constant fields which satisfy
$|\Phi|=|\bar\Phi|$ at vanishing chemical potential.  We review here
several forms and some basic features of the mean-field effective
potential for the Polyakov loop frequently used in the literature.
This effective potential will be incorporated in the thermodynamic
potential of the PQM model. The simplest effective potential is of a
Landau type, constructed with terms consistent with the $\mathbb{Z}_3$
symmetry \cite{pisarski00}:
\be
\beta^4\,{\cal U}_\tn{poly}(\Phi,\bar\Phi)=
-\frac{b_2(T)}{2}\Phi\bar\Phi -\frac{b_3}{6}(\Phi^3 + \bar\Phi^3)
+\frac{b_4}{4}(\Phi\bar\Phi)^2\, ,
\label{Eq:P_eff_pot_poly}
\ee  
where the temperature-dependent coefficient which makes
spontaneous symmetry breaking possible is 
\be
b_2(T)=a_0 + a_1\left(\frac{T_0}{T}\right) +a_2\left(\frac{T_0}{T}\right)^2 
+a_3 \left(\frac{T_0}{T}\right)^3\,.
\ee
$T_0$ is the transition temperature of the confinement/deconfinement
phase transition, in the pure gauge theory $T_0=270$~MeV.  The
parameters $a_i, i=0,\dots, 3$ and $b_3, b_4$ determined in
\cite{ratti06} reproduce the data measured in pure $SU(3)$ lattice
gauge theory \cite{boyd96} for pressure, and entropy and energy
densities.  The minimum of this potential is at $\Phi=0$ for low
temperature and $\Phi\to 1$ for $T\to \infty$ in accordance with the
definitions (\ref{Eq:Polyakov_op}) and
(\ref{Eq:Polyakov_field}). However, when using this potential in
either the PNJL or the PQM models the minimum of the resulting
thermodynamic potential is at $\Phi>1$ for $T\to\infty$ and also leads
to negative susceptibilities \cite{sasaki07}.

An effective potential for the Polyakov loop inspired by a
strong-coupling expansion of the lattice QCD action was derived in
\cite{fukushima04b,fukushima04a}. Using the part coming from the
$SU(3)$ Haar measure of group integration an effective potential was
constructed in \cite{ratti07,roessner07}
\be 
\beta^4\,{\cal U}_{\text{log}}(\Phi,\bar\Phi)= -\frac{1}{2}a(T) \Phi\bar \Phi 
+ b(T) \ln \left[1-6 \Phi\bar\Phi + 4\left(\Phi^{3}+\bar\Phi^{3}\right)
  - 3 \left(\Phi\bar\Phi\right)^{2}\right]\, ,
\label{Eq:P_eff_pot_log}
\ee
with the temperature-dependent coefficients
\be
a(T)=a_0+a_1 \left(\frac{T_0}{T}\right)+a_2 \left(\frac{T_0}{T}\right)^2,
\qquad
b(T)=b_3\left(\frac{T_0}{T}\right)^3\ .
\ee
The parameters $a_i,i=0,1,2$ and $b_3$ determined in \cite{ratti07} 
reproduce the thermodynamic quantities in the pure $SU(3)$ gauge theory
measured on the lattice. The use of this effective potential
cures the problem with negative susceptibilities \cite{sasaki07} 
and since the logarithm in ${\cal U}_{\text{log}}(\Phi,\bar\Phi)$ diverges 
as $\Phi,\bar\Phi\to 1$ it will also guarantee that 
$\Phi,\bar\Phi\to 1$ for $T\to \infty.$

A third effective potential is the one determined in 
Refs.~\cite{fukushima04b,fukushima04a}:
\be
\beta\, {\cal U}_{\tn{Fuku}}(\Phi,\bar\Phi) = 
-b \left[
54 e^{-a/T} \Phi \bar\Phi +\ln\left[1 - 6 \Phi \bar\Phi 
+ 4 (\Phi^3 + \bar\Phi^3)
- 3 (\Phi \bar\Phi)^2 \right]
\right],
\label{Eq:P_eff_pot_Fuku}
\ee
where $a$ controls the temperature of the deconfinement phase
transition in pure gauge theory, while $b$ controls the weight of 
gluonic effects in the transition. In this case, the parameters
$a=664$~MeV and $b=(196.2 \tn{MeV})^3$ are obtained from the
requirement of having a first order transition at about $T=270$~MeV
\cite{fukushima08,schaefer07}. 

It was shown in \cite{fukushima08} that there is little difference in
the pressure calculated from the three effective potentials for the
Polyakov loop in their validity region up to $T\simeq (1.5-2) T_c.$
The presence of dynamical quarks influences an effective treatment
based on the Polyakov loop which in this case is not an exact order
parameter.  Defining effective Polyakov-loop potentials for
nonvanishing chemical potential when $|\Phi|\ne|\bar\Phi|$ is
somewhat ambiguous \cite{schaefer07}.  In the present analysis we will
use at $\mu_q\ne 0$ the $\mathbb{Z}_3$ symmetric Polyakov-loop
potentials given above. The effect of the dynamical quarks was modeled by
considering the $N_f$ and $\mu_q$ dependence of the $T_0$ parameter 
of the Polyakov-loop effective potential. Using renormalization group 
arguments this dependence was parametrized in \cite{schaefer07} 
to be of the form $T_0(\mu_q,N_f)=T_\tau \exp(-1/(\alpha_0 b(\mu_q)))$. 
The parameters were chosen to have $T_0(\mu_q=0,N_f=2)=208.64$~MeV. When using 
the polynomial and logarithmic effective potential for the Polyakov loop
given in (\ref{Eq:P_eff_pot_poly}) and (\ref{Eq:P_eff_pot_log}) 
we will consider in addition to $T_0=270$~MeV two more cases, 
one with a constant value $T_0=208$~MeV and the other with the 
above-mentioned $\mu_q$-dependent $T_0$ taken at $N_f=2.$
 
\subsection{Constructing the grand potential of the PQM 
in a ``$\Phi$-derivable'' approximation}

The Lagrangian of the $SU(2)_L\times SU(2)_R$ chiral quark model
\cite{gell-mann60} is written in the usual matrix form 
\cite{bardeen68,roder03} by introducing two $N_f\times N_f$ matrices
\be
M(x)=\frac{\sigma(x)}{\sqrt{2N_f}}\un+i T^a \pi^a(x),\qquad
M_5(x)=\frac{\sigma(x)}{\sqrt{2N_f}}\un+i \gamma_5 T^a \pi^a(x),\qquad
\label{Eq:M_and_M5}
\ee
in terms of which one has
\bea
{\cal L}&=&\tr_f\left[\partial_\mu M^\dagger\partial^\mu M-m^2 M^\dagger M \right]
-\frac{\lambda}{6 N}\left[\tr_f\left(M^\dagger M\right)\right]^2 + N_f h\sigma
+\bar\psi\left(i\gamma^\mu D_\mu-\tilde g M_5\right)\psi +\tn{c.t.}
\,,
\label{Eq:Lagrange_matrix}
\eea
where in the mesonic part we have introduced an explicit symmetry
breaking term and ``c.t.''  stands for counterterms. 
After performing the trace, one can see that without the fermionic term
the Lagrangian (\ref{Eq:Lagrange_matrix}) is that of the $O(N)$ model, 
which describes the system of sigma and $N-1$ pion fields and is appropriately 
parametrized for a large-$N$ treatment \cite{patkos02}. Vanishing background 
is considered for the spatial components of the gauge 
field and a constant mean-field $A_0$ for the temporal component, so that 
the covariant derivative is $D_\mu=\partial_\mu-i \delta_{\mu 0} A_0.$ 
The trace in (\ref{Eq:Lagrange_matrix}) is to be taken in the flavor
space and to simplify notations the flavor, color, and Dirac indices
of the fermionic fields $\psi,\bar\psi$ are not indicated. The
$SU(N_f)$ generators in the fundamental representation $T^a$
($a=1,\dots, N_f^2-1$) satisfy the normalization condition 
$\tr_f(T^a T^b)=\delta^{ab}/2.$ Some rescaling with $N_f=\sqrt{N}$ 
was done and since in the mesonic sector we only want to increase 
the number of pions we do not introduce another invariant, 
$\tr_f\big[\big(M^\dagger M\big)^2\big],$ which for $N_f>2$ 
is independent of $\big[\tr_f\big(M^\dagger M\big)\big]^2.$ For a
recent treatment of the $U(N_f)_L\times U(N_f)_R$ meson model having both
invariants see \cite{fejos10}.

The constituent quarks become massive only after spontaneous/explicit
symmetry breaking. In (\ref{Eq:Lagrange_matrix}) the sigma field is
shifted by the vacuum expectation value $v$ as $\sigma\to v\sqrt{N}+\sigma,$
where on the right-hand side of the arrow $\sigma$ denotes the
fluctuating part of the original field. Then evaluating the trace, 
one obtains
\bea
{\cal L}&=&
-N\left[\frac{\lambda}{24} v^4+\frac{1}{2}m^2v^2 -h v\right]-
\sqrt{N}\left[\frac{\lambda}{6} v^3+m^2 v - h\right]\sigma
\nonumber\\
&&
+\frac{1}{2}\bigl[(\partial\sigma)^2 + (\partial\vec\pi)^2 \bigr]
-\frac{1}{2}m^2_{\sigma 0}\sigma^2
-\frac{1}{2}m^2_{\pi 0}\vec\pi^2
-\frac{\lambda v}{6\sqrt{N}}\sigma \rho^2-
\frac{\lambda}{24N}\rho^4 
\nonumber\\
&&+\bar \psi\big[(i\partial^\mu+\delta^{\mu0}A_0)\gamma_\mu-m_q\big]\psi
-\frac{g}{\sqrt{N}}\left[\bar\psi
\left(\sigma +i\sqrt{2N_f}\gamma_5T^a\pi^a\right)\right]\psi+
\tn{c.t.},
\label{Eq:Lagrange_field}
\eea  
where $\rho^2=\sigma^2+\vec\pi^2.$  
Here a rescaled Yukawa coupling $g=\tilde g \sqrt{N/(2N_f)}$ was 
introduced in order to assure the finiteness of the tree-level quark 
mass $m_q=g v$ in the $N\to\infty$ limit. In this limit, 
due to the $N_f$ scaling of the vacuum expectation value of the 
sigma field in (\ref{Eq:M_and_M5}) and 
of the coupling $\lambda$ in (\ref{Eq:Lagrange_matrix}) the 
tree-level sigma and pion masses $m^2_{\sigma 0}=m^2+\lambda v^2/2$ 
and $m^2_{\pi 0}=m^2+\lambda v^2/6$ are also finite.

After continuation to imaginary time, the grand partition function $Z$ 
and the grand potential $\Omega(T,\mu_B)$ of the spatially uniform 
system defined by (\ref{Eq:Lagrange_field}) are introduced as follows:
\be
Z=\tr\big\{
\exp\big[-\beta\big( H_0(A_4)+H_\tn{int}-\mu_B Q_B\big)\big]
\big\}=e^{-\beta \Omega},
\label{Eq:Z_def}
\ee
where $\mu_B$ is the baryon chemical potential. $H_\tn{int}$ is the 
interacting part of the Hamiltonian constructed from (\ref{Eq:Lagrange_field}).
With the help of $H_0$, the quadratic part of the Hamiltonian at vanishing 
$A_4,$ one can define the $A_4$-dependent free Hamiltonian $H_0(A_4),$ 
which for two quark flavors $u$ and $d$ reads as
\be
H_0(A_4)=H_0+\int d^3 x \left[i u^\dagger_i(x) A_{4,ij} u_j(x) + 
i d^\dagger_i(x) A_{4,ij} d_j(x)\right],
\ee
where $A_4=\tn{diag}(\phi_+,\phi_-,-(\phi_++\phi_-))$ is diagonal in color 
space and $i,j$ denotes color indices. In (\ref{Eq:Z_def}) $Q_B$ is
the conserved baryon charge which can be written in terms of the particle
number operators as $Q_B=\frac{1}{3}\sum\limits_{i=1}^3 N_{q,i},$ with
$N_\tn{q,i}=N_{u,i}+N_{d,i}-N_{\bar u,i}-N_{\bar d,i}$ and {\it e.g. } 
$N_{u,i}=\int d^3 x \big(u_i^\dagger u_i+d_i^\dagger d_i).$
Then combining $H_0(A_4)$ and $\mu_B Q_B$ one can see that the effect of 
fermions propagating on the constant background $A_4$, diagonal in 
the color space, is like having imaginary chemical potential for color.
Following Ref.~\cite{korthals_altes00} one introduces 
color-dependent chemical potential for fermions
\be
\mu_{1,2}=\mu_q-i\phi_\pm,\quad 
\mu_3=\mu_q+i(\phi_++\phi_-),
\label{Eq:c-dep_mu}
\ee
where $\mu_q=\mu_B/3$ is the quark baryon chemical potential.
Then, introducing the notation 
${\cal H}=H_0- \sum\limits_{i=1}^3\mu_i N_{q,i},$ 
one can write $Z$ as a path integral over the fields, generically 
denoted by~$\Psi$ 
\be
Z=e^{-\beta \Omega_0}
\frac{\displaystyle
\int\big[{\cal D}\Psi\big] \bigg\{
e^{-\beta {\cal H}} 
{\cal P} \exp\Big[-\int_0^\beta d\tau H_\tn{int}(\tau)\Big] \bigg\}
}
{\displaystyle \int\big[{\cal D}\Psi\big] 
e^{-\beta {\cal H}}},
\ee
where $\Omega_0$ is the grand potential of the
unperturbed system with fermions having color-dependent chemical potential 
\be
e^{-\beta \Omega_0}=
\int\big[{\cal D}\Psi\big] e^{-\beta {\cal H}}.
\ee

In the one-particle irreducible (1PI) formalism the grand potential is
given by the sum of the grand potential of the unperturbed system and
of perturbative quantum corrections represented by closed loops
constructed with the tree-level (free) propagators. In the
``$\Phi$-derivable'' approximation of Ref.~\cite{ward60}, also called
two-particle irreducible (2PI) approximation, the grand potential is a 
functional of the full propagators and field expectation values, and is 
of the following form:
\bea
\beta\Omega[G_\pi,G_\sigma,G,v,\Phi,\bar\Phi]&=&U(\Phi,\bar\Phi)+
\frac{N}{2} m^2 v^2+N\frac{\lambda}{24} v^4 - N h v
-(N-1)\frac{i}{2}\int_k 
\left[\ln G_\pi^{-1}(k)+D^{-1}_\pi(k) G_\pi(k)\right]
\nonumber\\
&&-
\frac{i}{2}\int_k\left[
\ln G_\sigma^{-1}(k)+D^{-1}_\sigma(k) G_\sigma(k)\right]
+\sqrt{N} i\, \tr_{D,c}\int_k 
\left[\ln G^{-1}(k)+D^{-1}(k) G(k)\right]
\nonumber\\
&&+
\Gamma_\tn{2PI}[G_\pi,G_\sigma,G,v,\Phi,\bar\Phi]+\tn{c.t.}\, ,
\label{Eq:Omega_grand_pot}
\eea
where the trace is taken in Dirac and color space. $U(\Phi,\bar\Phi)$ 
is a particular version of the effective Polyakov-loop potential 
reviewed in Sec.~\ref{ss:PEP}; the tree-level propagators of 
the pion, sigma, and constituent quark fields are 
\be
i D_\pi^{-1}(k)=k^2-m^2_{\pi0},\qquad
i D_\sigma^{-1}(k)=k^2-m^2_{\sigma0},\qquad
i D^{-1}(k)=\slk-m_q,\qquad
\ee
while $G_\pi,$ $G_\sigma,$ and $G$ are the respective full propagators.
$\Gamma_\tn{2PI}[G_\pi,G_\sigma,G,v,\Phi,\bar\Phi]$ denotes the set
of 2PI skeleton diagrams constructed with full
propagators, which to ${\cal O}(1/\sqrt{N})$ accuracy is given by
\bea
\Gamma_\tn{2PI}[G_\pi,G_\sigma,G,v,\Phi,\bar\Phi]
&=&
N\frac{\lambda}{24}\left(\int_k G_\pi(k)\right)^2
+\frac{\lambda}{12} \int_k G_\pi(k) \int_p G_\sigma(p)
-\frac{\lambda}{12}i\int_k \Pi(k) 
-\frac{i}{2}\int_k\ln\bigg(1-\frac{\lambda}{6}\Pi(k)\bigg)
\nonumber\\
&&
-\frac{\lambda}{6} v^2\int_k G_\sigma(k)
+\frac{\lambda}{6} v^2\int_k \frac{G_\sigma(k)}{1-\lambda \Pi(k)/6}
\nonumber\\
&&
-\sqrt{N}\frac{g^2}{2}i\tr_{D,c}\int_k\int_p\gamma_5 G(k)\gamma_5 G(k+p)G_\pi(p)
+\frac{g^2}{2\sqrt{N}}i\tr_{D,c}\int_k\int_p G(k) G(k+p) G_\sigma(p)\,, \ \ \
\label{Eq:Omega_2PI}
\eea
where the notation $\displaystyle \Pi(k)=-i\int_p G_\pi(p) G_\pi(k+p)$
was introduced. The mesonic part of $\Gamma_\tn{2PI}$ contains the 2PI
diagrams of the $O(N)$ model as given in Eq.~(2.13) and Figs.~2 and 4 of
\cite{dominici93} and also in Eq.~(48) and Fig.~2 of \cite{fejos09}.
We see that the contribution of the fermions which goes with
fractional powers of $N$ intercalates between the leading order (LO)
and next-to-leading order (NLO) contributions of the pions, which go
with integer powers of $N.$ This can be also seen by comparing the
expression of the pion propagator derived in (\ref{Eq:Gp}) with
Eqs.~(53) and (54) of \cite{fejos09}.

We use the imaginary time formalism of the finite 
temperature field theory in which the four-momentum is $k=(i\omega_n,\k).$ 
The Matsubara frequencies are $\omega_n=2\pi n T$ for bosons
while for fermions they depend also on the color due to the color-dependent 
chemical potential $\mu_i$ introduced in (\ref{Eq:c-dep_mu}) and are given by
$\omega_n=(2 n+1)\pi T-i \mu_i.$ The meaning of the integration symbol 
in (\ref{Eq:Omega_2PI}) is
\be
\int_k = i T\sum_{n} \int_{\k}\equiv i T\sum_{n}\int\frac{d^3 \k}{(2\pi)^3}.
\label{Eq:sum_int_def}
\ee
The dependence on $\Phi$ and $\bar\Phi$ of the fermionic trace-log term 
in the grand potential $\Omega$ and of the quark-pion setting-sun in 
$\Gamma_\tn{2PI}$ results from the fact that, as shown in the Appendix 
between Eqs.~(\ref{Eq:I1_def}) and (\ref{Eq:I1_final}), as well as between  
Eqs.~(\ref{Eq:I2}) and (\ref{Eq:T2_decompose}), respectively,
after performing the Matsubara sum the color trace can be 
expressed in closed form in terms of the mean-field ($\vec x$-independent) 
Polyakov loop $\Phi$ and its conjugate $\bar \Phi.$

What we evaluate in this work is not the grand potential 
(\ref{Eq:Omega_2PI}), but rather its 
derivatives, that is the equations for the two-point functions and the 
field equations, which are given by the stationary conditions
\be
\frac{\delta \Omega}{\delta G}=\frac{\delta \Omega}{\delta G_\pi}=
\frac{\delta \Omega}{\delta G_\sigma}=\frac{\delta \Omega}{\delta v}=
\frac{\delta \Omega}{\delta \Phi}=\frac{\delta \Omega}{\delta \bar\Phi}=0.
\label{Eq:stac_cond}
\ee 
In each of these equations we will keep the contribution of the
fermions only at the leading order in the large-$N_f$ expansion. The LO
contribution of the fermions is ${\cal O}(\sqrt{N})$ in the field
equations of $\Phi$ and $\bar\Phi,$ ${\cal O}(1)$ in the equation for
the fermion propagator $G,$ and ${\cal O}(1/\sqrt{N})$ in the
remaining equations, that is the field equation of $v$, and the equations
of $G_\pi$ and $G_\sigma.$

It is easy to see that the third and fourth terms on the right-hand side of 
(\ref{Eq:Omega_2PI}) do not contribute to any of the equations at the
order of interest, and that the second term contributes
only in the equation for the sigma propagator 
\be
i G_\sigma^{-1}(p)=i D_\sigma^{-1}(p)+\frac{\lambda v^2}{3}-
\frac{\lambda}{6}\int_k G_\pi(k)-
\frac{\lambda v^2}{3} \frac{1}{1-\lambda \Pi(p)/6}
-\frac{i g^2}{\sqrt{N}}\tr_{D,c}\int_k G(k) G(k+p) + \tn{c.t.}.
\label{Eq:Gs}
\ee
In fact, the equation for $G_\sigma$ decouples, in the sense that $G_\sigma$
will not appear in any of the remaining five equations. Nevertheless,
it plays an important role in the parametrization of the model, 
as will be shown in Sec.~\ref{ss:param}.

\subsection{Approximations made to solve the model \label{ss:approx}}

In this work we use some approximations to solve the set of 
coupled equations coming from (\ref{Eq:stac_cond}). 

1.~As a first approximation we disregard the self-consistent equation for 
the fermions arising from $\delta\Omega/\delta G=0,$ that is
\be
i G^{-1}(k)=i D^{-1}(k)-i g^2\int_p \gamma_5 G(p) \gamma_5 G_\pi(p-k) 
+ \tn{c.t.},
\label{Eq:G}
\ee 
and simply use  in the remaining five equations the tree-level fermion
propagator $D(k).$  A study based on the solution
of the self-consistent equation for the fermion propagator is 
left for a forthcoming publication.

Within this approximation the field equation for $v$, hereinafter
called equation of state (EoS), and the pion propagator simplify
considerably. The contribution of the last but one term of
(\ref{Eq:Omega_2PI}) to the pion propagator breaks up upon working out
the Dirac structure into the linear combination of a fermionic tadpole
$\tilde T(m_q)$ and a bubble integral $\tilde I(p;m_q)$. 
Introducing the propagator
\be
D_0(k)=\frac{i}{k^2-m_q^2},
\label{Eq:D0_prop}
\ee
these integrals are defined as
\bea
\label{Eq:T_q_def}
\tilde T(m_q)&=&\frac{1}{N_c}\sum_{i=1}^{N_c}\int_k D_0(k),\\
\tilde I(p;m_q)&=&\frac{1}{N_c}\sum_{i=1}^{N_c}
\left[-i\int_q D_0(q) D_0(q+p)\right].
\label{Eq:I_q_def}
\eea
In terms of these integrals which are evaluated in the Appendix between 
Eqs.~(\ref{Eq:T_q}) and (\ref{Eq:I_q_beta_finite}) one obtains:
\bea
0&=&N v\left[m^2+\frac{\lambda}{6}\left(v^2+\int_k G_\pi(k)\right)
-\frac{4 g^2 N_c}{\sqrt{N}}\tilde T(m_q)-\frac{h}{v} 
\right] + \tn{c.t.}\, ,
\label{Eq:EoS}
\\
i G_\pi^{-1}(k)&=&k^2-m^2-\frac{\lambda}{6}\left(v^2+\int_k G_\pi(k)\right)
+ \frac{4g^2 N_c}{\sqrt{N}} \tilde T(m_q) - 
\frac{2 g^2 N_c}{\sqrt{N}} k^2 \tilde I(p;m_q) + \tn{c.t.}\, .
\label{Eq:Gp}
\eea  
One can see that the Goldstone theorem is fulfilled, since using the
EoS in the equation for the pion propagator one obtains 
$i G_\pi^{-1}(k=0)=-h/v.$ This is only accidental because the Ward identity
relating the inverse fermion propagator and the proper vertex
$\Gamma_{\pi^a\psi\bar\psi}=\delta^3 \Gamma/\delta \bar\psi\delta \psi \delta \pi^a$ (see {\it e.g.} Eq.~(13.102) of \cite{zinn-justin})
\be
-\frac{i}{2} T_a\Big\{\gamma_5,i G^{-1}(p)\Big\}=v\sqrt{\frac{N}{2 N_f}}
\Gamma_{\pi^a\psi\bar\psi}(0,p,-p),
\label{Eq:WI}
\ee
is satisfied only with tree-level propagators and vertices. The
relation above is violated at any order of the perturbation theory in
the large-$N_f$ approximation, since in view of (\ref{Eq:G}) the
corrections to the inverse tree-level fermion propagator are of 
${\cal O}(1),$ while the corrections to the tree-level 
$\pi-\psi-\bar\psi$ vertex are suppressed by $1/N.$

2.~A further approximation concerns the self-consistent pion
propagator (\ref{Eq:Gp}). In this work four approximations for $G_{\pi}$
are considered; two local approximations and two nonlocal approximations
obtained using an expansion in $1/\sqrt{N}.$ In the local
approximation one parametrizes the pion propagator as
\be
G_{\pi,l}(p)=\frac{i}{p^2-M^2},
\label{Eq:Gp_local}
\ee
and uses this form in all equations. In the {\it first} variant 
$M^2$ is determined as a pole mass from 
$i G^{-1}_{\pi,l}(p_0^2=M^2,\p=0)=0$ by the self-consistent gap equation
arising from (\ref{Eq:Gp})
\be
M^2=m^2+\frac{\lambda}{6} \left(v^2+T_F(M)\right)
-\frac{4 g^2 N_c}{\sqrt{N}}\tilde T_F(m_q)
+\frac{2 g^2 N_c}{\sqrt{N}}M^2 \tilde I_F(M,\bm{0};m_q).
\label{Eq:gap_pole}
\ee
In a {\it second} variant $M^2$ is determined from 
$M^2=-i G^{-1}_{\pi,l}(p=0)$, when the gap-equation becomes
\be
M^2=m^2+\frac{\lambda}{6}\left(v^2+T_F(M)\right)
-\frac{4g^2 N_c}{\sqrt{N}} \tilde T_F(m_q).
\label{Eq:gap_p0}
\ee
The subscript $F$ denotes the finite part of the integrals defined in
Eqs.~(\ref{Eq:T_q_def}) and (\ref{Eq:I_q_def}), which are given
explicitly in Eq.~(\ref{Eq:Tad_q_F_decomp}) and
Eqs.~(\ref{Eq:I_q_finite})-(\ref{Eq:I_q_beta_finite}).  In this way
the finite parts of all vacuum pieces are contained in our
equation. The importance of these terms for the thermodynamics of the
PQM model was pointed out in \cite{nakano10}. In view of the EoS
(\ref{Eq:EoS}) the two definitions of $M^2$ coincide in the chiral
limit $h=0,$ where for both variants one has $M^2=0.$ We note that due
to their self-consistent nature, when (\ref{Eq:gap_pole}) or
(\ref{Eq:gap_p0}) is solved, a series containing all orders of
$1/\sqrt{N}$ is in fact resummed.

The {\it third}, nonlocal variant of the pion equation is 
derived using an $1/\sqrt{N}$ expansion in the pion propagator (\ref{Eq:Gp})
after exploiting the EoS (\ref{Eq:EoS}). One obtains
\bea
G_\pi(p)&=&\frac{i}{p^2-\frac{h}{v}-\frac{2 g^2 N_c}{\sqrt{N}}p^2 
\tilde I_F(p;m_q)}
=\frac{i}{p^2-\frac{h}{v}}+\frac{2 g^2 N_c}{\sqrt{N}} 
\frac{i p^2 \tilde I_F(p;m_q)}{\left(p^2-\frac{h}{v}\right)^2}
+{\cal O}\left(\frac{1}{N}\right).
\label{Eq:Gp_third}
\eea  
With this form of the pion propagator the EoS reads
\be
m^2+\frac{\lambda}{6}\left(v^2+T_F(M)\right)
+\frac{2 g^2 N_c}{\sqrt{N}} J_F(M,m_q)
-\frac{4 g^2 N_c}{\sqrt{N}} \tilde T_F(m_q)=\frac{h}{v},
\label{Eq:EoS_third}
\ee
where in this case $M^2=h/v$ and we have introduced the integral
\be
J(M,m_q)=-i\int_p G^2_{\pi,l}(p) p^2 \tilde I_F(p;m_q).
\label{Eq:J_def}
\ee
Solving this equation for $v$ shows that this approximation still resums 
infinitely many orders in $1/\sqrt{N}$.

A {\it fourth} variant of $G_{\pi},$ which by a strict expansion
in $1/\sqrt{N}$ will include terms of no higher order than 
${\cal O}(1/\sqrt{N}),$ can be obtained by expanding not only the
nonlocal, momentum-dependent part of the self-energy in the pion
propagator (\ref{Eq:Gp}), but also its local part. This is explicitly
constructed including counterterms in Sec.~\ref{sec:renorm}, where a
diagrammatic illustration of the approximation is also given. 
For this approximation, the pion propagator is given by 
Eqs.~(\ref{Eq:Gp_expanded}), (\ref{Eq:M2_LO_finite}), 
and (\ref{Eq:M2_NLO_finite}), while the EoS is given in (\ref{Eq:d_all}).

\subsection{Parametrization\label{ss:param}}

The mass parameter $m^2,$ the couplings $g,\lambda,$ the
renormalization scale $M_{0B},$ the vacuum expectation value $v_0$,
and the external field $h,$ which vanishes in the chiral limit, are
determined at $T=\mu_q=0$ using some information from the sigma sector
and the following physical quantities: the pion decay constant
$f_\pi=93$~MeV and its mass $m_\pi=140$~MeV, and the constituent quark
mass taken to be $M_q=m_N/3=313$~MeV. From the sigma sector we use the
mass and the width of the sigma particle and the behavior of the
spectral function.  $v_0$ is determined from
the matrix element of the axial vector current between the vacuum
state and a one-pion state, which due to the rescaling of the vacuum
expectation value by $\sqrt{N}$ gives $v_0=f_\pi/2.$ The value of the
Yukawa coupling $g=6.7$ is obtained by equating the tree-level fermion
mass $m_q$ with $M_q.$ The parameters $\lambda$ and $M_{0B}$ are
determined from the sigma propagator, as will be detailed below.
Having determined them, in the chiral limit $m^2$ is fixed from the EoS,
while in the case of the physical pion mass the remaining parameters $m^2$
and $h$ are determined as follows. If the local approximation for the
pion propagator is used $m^2$ is determined from the gap equation by
requiring $M^2=m_\pi^2,$ and $h$ is obtained from the EoS.  When
$G_\pi$ is approximated using a large-$N_f$ expansion $h$ is fixed by
requiring $h=m_\pi^2 v_0,$ and $m^2$ is determined from the EoS.

Now we turn to the issue of fixing $\lambda$ and $M_{0B}.$
Using in (\ref{Eq:Gs}) the tree-level fermion propagator together with
the local approximation (\ref{Eq:Gp_local}) for the pion propagator
and also the equation of state (\ref{Eq:EoS}), one obtains after working out 
the Dirac structure the following form for the sigma propagator:
\be 
i G_\sigma^{-1}(p)=p^2-\frac{h}{v}-\frac{\lambda v^2}{3}
\frac{1}{1-\lambda I_F(p;M)/6}+\frac{2 g^2 N_c}{\sqrt{N}} (4m_q^2-p^2)
\tilde I_F(p;m_q).
\label{Eq:Gs_param}
\ee
The integral $I_F(p;M),$ obtained using the local approximation 
(\ref{Eq:Gp_local}) for the pion propagator with $M^2=m_\pi^2,$
 can be found in Eqs.~(10) and (11) of Ref.~\cite{patkos02} with 
$M_0$ replaced by $M_{0B},$ while $\tilde I_F(p;m_q)$ is given in 
Eqs.~(\ref{Eq:I_q_finite})-(\ref{Eq:I_q_beta_finite}).

\begin{figure}[t]
\begin{center}
\includegraphics[keepaspectratio,width=0.5\textwidth,angle=0]{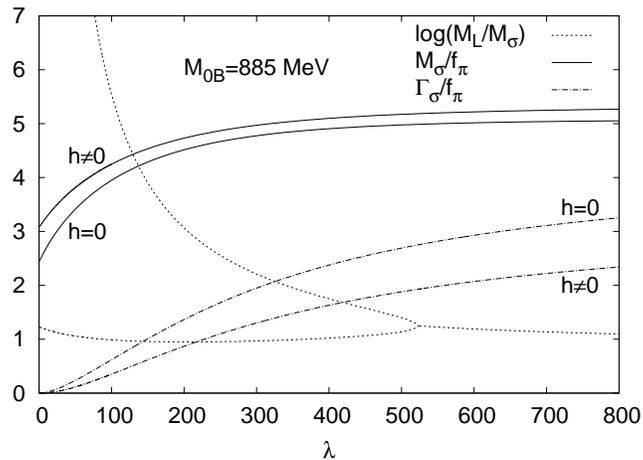}
\caption{The $\lambda$ dependence of the real and imaginary parts of the
complex sigma pole $p_0=M_\sigma-i\Gamma_\sigma/2$ and of the Landau ghost 
$M_L$ in the chiral limit and for the physical pion mass. The upper curve for 
$M_\sigma$ and the lower one for $\Gamma_\sigma$ represent the case of 
the physical pion mass, as indicated on the plot. $M_L$ is shown only 
in this case, for in the chiral limit there is very little difference.}
\label{Fig:sigma}
\end{center}
\end{figure}

Both in the chiral limit $M=0$ and for $M=m_\pi$ the self-energy has
along the positive real axis of the complex $p_0$ plane two cuts above
the thresholds of the pion and fermion bubble integrals, which start
at $p^2=4 M^2$ and $p^2=4 m_q^2,$ respectively. Above these thresholds
the respective pion and fermion bubble integrals have nonvanishing
imaginary parts.  We search for poles of the sigma propagator
analytically continued between the two cuts to the second Riemann
sheet in the form $i G^{-1}_\sigma(p_0=\kappa e^{-i\phi},\p=0)=0$.
The pole is parametrized as $p_0=M_\sigma-i \Gamma_\sigma/2,$ with the
real and imaginary parts corresponding to the mass and the half-width
of the sigma particle.  The solution for $M_\sigma$ and
$\Gamma_\sigma$ is shown in Fig.~\ref{Fig:sigma} both in the chiral
limit ($h=m_\pi=0$) and for the $h\ne 0$ case.  Similar to the case of
the $O(N)$ model studied in Ref.~\cite{patkos02}, in the chiral limit
the value of $M_\sigma$ is a little smaller and the value of
$\Gamma_\sigma$ larger than in the $h\ne 0$ case.  Comparing
Fig.~\ref{Fig:sigma} with Fig.~2 of Ref.~\cite{patkos02} obtained in
the $O(N)$ model, that is without fermions, the $M_\sigma(\lambda)$
curve moved slightly upward, but the $\Gamma_\sigma(\lambda)$ curve
moved significantly downward, which means that in the present case the
phenomenologically expected value \cite{caprini06}
$M_\sigma/\Gamma_\sigma\sim 1$ cannot be achieved for any value of the
coupling $\lambda.$ Another difference is that for low values of
$\lambda$ there are two poles of $G_\sigma$ on the negative imaginary
axis in contrast to only one such pole in the $O(N)$ model. These
poles approach each other as $\lambda$ increases and after they
collide at a given value of $\lambda$ there are two complex poles at
higher $\lambda$, one with positive and one with negative real
part. The imaginary part of the complex pole having positive real part
is shown in Fig.~\ref{Fig:sigma} for the renormalization scale
$M_{0B}=885$~MeV. As explained in the study done in the chiral limit
in \cite{toni04} for lower values of the renormalization scale the
scale $M_L$ of the lower Landau ghost on the imaginary axis comes even
closer to $M_\sigma$ and as a result the spectral function of the
sigma is heavily distorted. In order to avoid this and based on the
ratio of $M_\sigma/\Gamma_\sigma$ we have chosen $\lambda=400$ and
$M_{0B}=885$~MeV. For these values $M_\sigma=456$~MeV and
$\Gamma_\sigma=221$~MeV in the chiral case, while $M_\sigma=474$~MeV
and $\Gamma_\sigma=152$~MeV for the case of a physical pion
mass. These values are used throughout Sec.~IV also in the case when
the pion propagator is expanded in $1/\sqrt{N},$ that is in the third
and fourth variants of $G_\pi$ discussed in
Sec.~\ref{ss:approx}.

\section{Renormalization \label{sec:renorm}}

The lesson one can learn from the successful renormalization in 2PI
\cite{hess02,blaizot04,borsanyi05,reinosa06,patkos08} or large-$N$
\cite{fejos09,toni09,fejos08} approximations is that, due to the
systematic nature of these expansions, the counterterms can be obtained
by analyzing the structure of the equations. In these cases there is
no need for an order-by-order detailed study of the counterterm
diagrams which becomes rather complicated due to the proliferation of
the diagrams.  If one uses some ad-hoc approximation which spoils the
self-consistent nature of the propagator equations or is not
systematic, then one loses the possibility to explicitly or uniquely
determine the counterterms from the equations.
 
In this section we discuss the renormalization of the model in the
case of a strict $1/\sqrt{N}$ expansion in the pion propagator and the
equation of state. This expansion is not entirely consistent because
as mentioned in Sec.~\ref{ss:approx} we use for simplicity
tree-level fermion propagators despite the fact that this expansion
produces ${\cal O}(1)$ corrections in the fermion propagator
(\ref{Eq:G}) which all have to be resummed. We will see that as a
consequence of this approximation the best one can achieve is to
determine the counterterms to some order in the Yukawa coupling. It
will turn out that the order depends on the equation and the type of
subdivergence we are looking at.

The counterterm functional needed to renormalize the pion propagator 
and the equation of state reads
\bea
\nonumber
\Omega_\tn{ct}[G_\pi,v]&=&
\frac{N}{2}\delta m_0^2 v^2+\frac{N}{24}\delta\lambda_4 v^4
+\frac{N-1}{2}\left(\delta m_2^2+\frac{\delta\lambda_2}{6} v^2\right)
\int_k G_\pi(k) \\
&&+(N-1)\frac{\delta\lambda_0}{24}\left[\int_k G_\pi(k)\right]^2
-(N-1)\frac{\delta Z}{2}\int_k k^2 G_\pi(k).
\label{Eq:Omega_ct}
\eea
Compared to the counterterm functional used in Eq.~(48) of
\cite{fejos09} to renormalize the stationary equations of the $O(N)$ 
model the only difference in (\ref{Eq:Omega_ct}) is the appearance of 
the term containing the wave-function renormalization counterterm 
$\delta Z.$ This is needed to remove the momentum-dependent divergence 
of the fermionic contribution to the pion propagator (\ref{Eq:Gp}) 
rewritten as
\be
i G_\pi^{-1}(k)=(1+\delta Z)k^2-M^2-\frac{2 g^2 N_c}{\sqrt{N}}\,
k^2 \tilde I(k;m_q).
\label{Eq:Gp_ct}
\ee
The local part $M^2$ containing the remaining counterterms reads
\be
M^2=m^2+\delta m_2^2+\frac{\lambda+\delta\lambda_2}{6} v^2
+\frac{\lambda+\delta\lambda_0}{6} \int_p G_\pi(p)
-\frac{4 g^2 N_c}{\sqrt{N}}\tilde T(m_q).
\label{Eq:M2_ct}
\ee
Using in (\ref{Eq:Gp_ct}) the notation introduced in (\ref{Eq:I_q_div}) 
one can readily determine $\delta Z:$
\be
\delta Z=\frac{2 g^2 N_c}{\sqrt{N}} \tilde I_\tn{div}(k;m_q)
=\frac{2 g^2 N_c}{\sqrt{N}} T_d^{(0)}.
\ee

Separating the LO and NLO contributions in the local part given in
(\ref{Eq:M2_ct}) by writing $M^2=M^2_\tn{LO}+M^2_\tn{NLO}/\sqrt{N},$ 
an expansion in powers of $1/\sqrt{N}$ in the pion propagator 
(\ref{Eq:Gp_ct}) gives
\bea
G_\pi(p)
=G_\tn{LO}(p)-
i \frac{G^2_\tn{LO}(p)}{\sqrt{N}}\left[
M^2_\tn{NLO}-2 N_c g^2 p^2 \tilde I_F(p;m_q)\right]
+{\cal O}\left(\frac{1}{N}\right),
\label{Eq:Gp_expanded}
\eea
where $G_\tn{LO}(p)=i/(p^2-M^2_\tn{LO}).$
The counterterms are also written as the sum of LO and NLO
contributions $\delta m_i^2=\delta {m^2_i}^{(0)}+\delta {m^2_i}^{(1)}/\sqrt{N},$
$\delta\lambda_i=\delta\lambda_i^{(0)}+\delta\lambda_i^{(1)}/\sqrt{N}$
and used together with (\ref{Eq:Gp_expanded}) in (\ref{Eq:M2_ct})
to obtain the equations for the LO and NLO local parts 
\bse
\bea
&&
M^2_\tn{LO}=m^2+\delta {m^2_2}^{(0)}+
\frac{\lambda+\delta\lambda_2^{(0)}}{6} v^2+
\frac{\lambda+\delta\lambda_0^{(0)}}{6} T(M_\tn{LO}),
\label{Eq:M2_LO}
\\
&&
M^2_\tn{NLO}
\left[
\frac{1}{\lambda_B^{(0)}}-\frac{I(0;M_\tn{LO})}{6}\right]=
\frac{1}{\lambda_B^{(0)}}\left[
\delta {m^2_2}^{(1)}+\frac{\delta\lambda_2^{(1)}}{6}v^2+
\frac{\delta\lambda_0^{(1)}}{6} T(M_\tn{LO})\right]
-\frac{4 g^2 }{\lambda_B^{(0)}} N_c \tilde T(m_q)
+\frac{2 g^2 }{6} N_c J(M_\tn{LO},m_q),\ \ \ 
\label{Eq:M2_NLO}
\eea
\ese
where we divided the second equation by 
$\lambda_B^{(0)}=\lambda+\delta\lambda_0^{(0)}$ and used the integral
introduced in (\ref{Eq:J_def}) with $G_{\pi,l}$ replaced by $G_\tn{LO}.$

Compared to the perturbative renormalization of the fermionic trace-log
contribution to the effective potential performed in \cite{skokov10b},
the difficulty here is that, due to the self-consistent nature of the
pion propagator, an infinite series of coupling counterterms of the mesonic 
part, that is $\delta\lambda_0,$ $\delta\lambda_2,$ $\delta m_0^2,$ and 
$\delta m_2^2$ has to be determined to ${\cal O}(\sqrt{N})$ of the 
large-$N$ expansion. In order to achieve this we apply the method developed 
in Refs. \cite{fejos08,patkos08,fejos10}. The method for determining the
counterterms appearing in a particular equation resides in the
separation of the divergent part of the integrals contained by the
equation. This is obtained by expanding the propagators around an
appropriately defined auxiliary propagator (see Appendix~A of
\cite{patkos08}).  Then, the finite part of the integrals is used to
write down a finite equation, which, when subtracted from the original
equation, provides a relation between counterterms and
divergences. This relation still involves the vacuum expectation value
$v$ and the finite part of some integrals, {\it e.g.} $T_F,$ the
finite part of the pion tadpole.  Requiring as in \cite{fejos08} the
vanishing of the coefficient of $v^2$ and $T_F$ (cancellation of
subdivergences) one obtains the coupling counterterms, while the
remaining part of the relation mentioned above gives the mass
counterterm (cancellation of the overall divergence).

To apply the above method to Eq.~(\ref{Eq:M2_LO}) for
$M^2_\tn{LO}$, which is the gap equation of the $O(N)$ model at
leading order of the large-$N$ approximation, one uses the expression
of the pion tadpole given in (\ref{Eq:Tad_pi}) in terms of finite and
divergent pieces. Retaining in (\ref{Eq:M2_LO}) the finite part of 
the tadpole one obtains the LO finite gap equation 
\be
M^2_\tn{LO}=m^2+\frac{\lambda}{6}\big[v^2+T_F(M_\tn{LO})\big].
\label{Eq:M2_LO_finite}
\ee
Subtracting this from  (\ref{Eq:M2_LO}) one requires the  vanishing of 
the coefficient of $v^2$ and $T_F(M_\tn{LO})$ in the resulting equation. This 
determines the LO coupling counterterms 
\be
\delta\lambda_2^{(0)}=\delta\lambda_0^{(0)}=
-\frac{\lambda^2}{6}\frac{T_d^{(0)}}{1+\lambda T_d^{(0)}/6},
\label{Eq:LO_delta_lambda}
\ee 
while requiring the cancellation of the remaining overall divergence 
determines the LO mass counterterm
$\delta {m^2_2}^{(0)}=-\big(\lambda+\delta\lambda_0^{(0)}\big)
\left[T_d^{(2)}+[M^2-M_0^2] T_d^{(0)}\right]/2.$ 

The determination of the counterterms in the equation for the NLO local part
in the pion propagator parallels to some extent the analysis of the
NLO divergences in the $O(N)$ model discussed in Sec.~VI~B
of \cite{fejos09}. One observes that since $I_\tn{div}(0;M_\tn{LO})=T_d^{(0)},$
in view of (\ref{Eq:LO_delta_lambda}) the left-hand side of 
(\ref{Eq:M2_NLO}) is finite and it actually enters the finite equation for
$M^2_\tn{NLO}$ 
\be
M^2_\tn{NLO}
\left[
\frac{1}{\lambda}-\frac{I_F(0;M_\tn{LO})}{6}
\right]=
-\frac{4 g^2}{\lambda}N_c \tilde T_F(m_q)
+\frac{2 g^2}{6} N_c J_F(M_\tn{LO},m_q).
\label{Eq:M2_NLO_finite}
\ee
Subtracting (\ref{Eq:M2_NLO_finite}) from (\ref{Eq:M2_NLO}) the following 
relation between divergences and counterterms is obtained:
\bea
0=\frac{1}{\lambda_B^{(0)}}\left[
\delta {m^2_2}^{(1)}+\frac{\delta\lambda_2^{(1)}}{6}v^2+
\frac{\delta\lambda_0^{(1)}}{6} T(M_\tn{LO})-4g^2\tilde T_\tn{div}(m_q)
\right]
-4 g^2 N_c \frac{T_d^{(0)}}{6}\tilde T_F(m_q)
+\frac{2 g^2}{6} N_c J_\tn{div}(M_\tn{LO},m_q).
\eea
Then we use the expression of $J_\tn{div}(M_\tn{LO},m_q)$ given in 
(\ref{Eq:J_div}) in terms of $M_\tn{LO},m_q^2,$ and $\tilde T(m_q)$ 
in which one substitutes for $M_\tn{LO}$ its expression from 
(\ref{Eq:M2_LO_finite}). The terms proportional to $\tilde T_F(m_q)$ 
cancel. The overall divergences determine the form of $\delta {m^2_2}^{(1)}.$
The remaining terms proportional to $v^2$ and $T_F(M_\tn{LO})$ 
can be grouped as
\bea
\nonumber
\dots
&+&
\frac{v^2}{6}\Bigg\{
\delta\lambda_2^{(1)}+
\frac{\lambda \delta\lambda_0^{(1)}}{6} T_d^{(0)}
+\frac{2 g^2}{3} N_c\lambda \lambda_B^{(0)} T_d^{(I)}
+4 g^4 N_c \left[\lambda_B^{(0)} 
\big(T_d^{(I)}+(T_d^{(0)})^2\big)-6 g^2 T_d^{(0)}
\right]
\Bigg\}
\\
&+&
\frac{T_F(M_\tn{LO})}{6}\,
\left[
\delta\lambda_0^{(1)}+
\frac{\lambda \delta\lambda_0^{(1)}}{6} T_d^{(0)}
+\frac{2 g^2}{3} N_c\lambda \lambda_B^{(0)} T_d^{(I)}
\right]=0.
\label{Eq:delta_lambda_NLO_Gp}
\eea
Requiring the vanishing of the coefficient of $v^2$ and $T_F(M_\tn{LO})$ 
determines the NLO coupling counterterms $\delta\lambda_2^{(1)}$ and
$\delta\lambda_0^{(1)}.$ One can see that these counterterms agree at
${\cal O}(g^2)$ but differ at ${\cal O}(g^4)$.

A completely similar analysis performed on the EoS 
\be
m^2+\delta m_0^2+\frac{\lambda+\delta\lambda_4}{6}v^2
+\frac{\lambda+\delta\lambda_2}{6}\int_k G_\pi(k) 
-\frac{4 g^2 N_c}{\sqrt{N}}\tilde T(m_q)=\frac{h}{v} 
\ee
gives an equation analogous in form and meaning 
to (\ref{Eq:delta_lambda_NLO_Gp})
\bea
\nonumber
\dots
&+&
\frac{v^2}{6}\,\Bigg\{
\delta\lambda_4^{(1)}+
\frac{\lambda \delta\lambda_2^{(1)}}{6} T_d^{(0)}
+\frac{2 g^2}{3} N_c \lambda \lambda_B^{(0)} T_d^{(I)}
+4 g^4 N_c \left[
\lambda_B^{(0)} \big(T_d^{(I)}+(T_d^{(0)})^2\big)-6 g^2 T_d^{(0)}\right]
\Bigg\}
\\
&+&
\frac{T_F(M_\tn{LO})}{6}
\left[\delta\lambda_2^{(1)}+
\frac{\lambda \delta\lambda_2^{(1)}}{6} T_d^{(0)}
+\frac{2 g^2}{3} N_c\lambda 
\Big(\lambda+\delta\lambda_0^{(0)}\Big) T_d^{(I)}
\right]=0.
\label{Eq:delta_lambda_NLO_EoS}
\eea

From this equation one can see that the NLO coupling counterterms
$\delta\lambda_2^{(1)}$ and $\delta\lambda_4^{(1)}$ agree at 
${\cal O}(g^2)$ but differ at ${\cal O}(g^4)$. Moreover, one sees that
$\delta\lambda_2^{(1)}$ determined from the requirement to cancel the
coefficient of $v^2$ in (\ref{Eq:delta_lambda_NLO_EoS}) differs at
${\cal O}(g^4)$ from $\delta\lambda_2^{(1)}$ needed to cancel the
coefficient of $T_F(M_\tn{LO})$ in (\ref{Eq:delta_lambda_NLO_EoS}). 
These requirements give the same expression for the counterterm only 
at ${\cal O}(g^2).$

The above feature is a consequence of the fact that by keeping the
fermions unresummed one does not take into account all the diagrams
which are of the same order in the large-$N_f$ expansion.  It does not
necessarily mean that the approximation we use is unrenormalizable.
Rather we suggest the interpretation that the approximation is such
that different subseries of the counterterms are needed to cancel the
subdivergences of different equations. Although we have a method to
determine the counterterms in each equation, corrections are to be
expected starting at ${\cal O}(g^4).$ If one traces back the origin of
the term proportional to $T_F(M_\tn{LO})$ in
(\ref{Eq:delta_lambda_NLO_Gp}) one finds that it comes from the
expression (\ref{Eq:M2_LO_finite}) for $M^2_\tn{LO}$ used in the
divergent contribution $J_\tn{div}(M_\tn{LO},m_q)$ as given by
(\ref{Eq:J_div}). In turn, the integral $J(M_\tn{LO},m_q)$ defined in
(\ref{Eq:J_def}) is generated through the expansion
(\ref{Eq:Gp_expanded}) by the one-loop fermion bubble contribution to
the pion self-energy. But, when the first correction to the fermion
propagator is included, then we have to take into account in the
square bracket of the expanded pion propagator (\ref{Eq:Gp_expanded})
the contribution of the two-loop self-energy
\be
i\raisebox{-0.41cm}{\includegraphics[width=1.75cm]{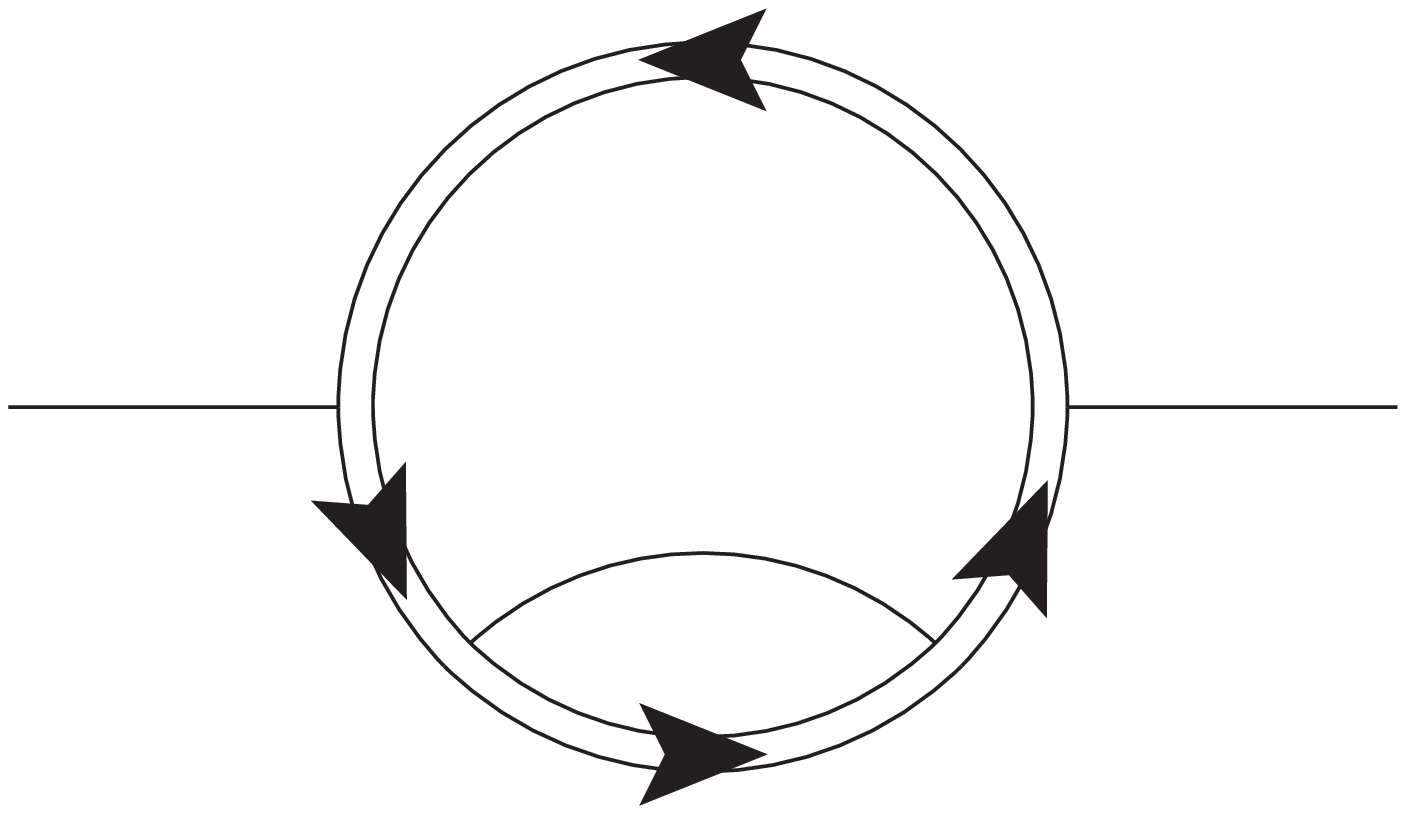}}=\frac{g^4  N_c}{\sqrt{N}} \Sigma_2(p;M_\tn{LO},m_q).
\label{Eq:Sigma2}
\ee
This will generate in the equation for $M^2_\tn{NLO}$ an integral similar to 
$J(M_\tn{LO},m_q)$
\be
g^4 K(M_\tn{LO},m_q)=-i g^4 \int_p G^2_\tn{LO}(p) \Sigma_2(p;M_\tn{LO},m_q),
\label{Eq:K}
\ee
which is expected to have a divergence proportional to
$T_F(M_\tn{LO}).$ This divergence would result in ${\cal O}(g^4)$
corrections in the $\delta\lambda_0^{(1)}$ counterterm as determined
from (\ref{Eq:delta_lambda_NLO_Gp}). Therefore, it is expected that
with a resummed fermion propagator, as required by the large-$N_f$
resummation, the determined counterterms will eventually agree in all
equations.  It is rather nontrivial to check this conjecture, even at
the two-loop level, because the reduction of the two-loop integral in
(\ref{Eq:Sigma2}) performed with the method of \cite{weiglein94}
produces more than a dozen scalar integrals and their contribution
should be analyzed in (\ref{Eq:K}). We have only checked that at
the two-loop level the Goldstone theorem is indeed violated as mentioned
in Sec.~\ref{ss:approx}, based on the violation of the Ward
identity (\ref{Eq:WI}).

\begin{figure}[htbp]
\centering
$\displaystyle i\sum_{\tn{loops}}$\raisebox{-0.4cm}{
\includegraphics[keepaspectratio,width=0.2\textwidth,angle=0]{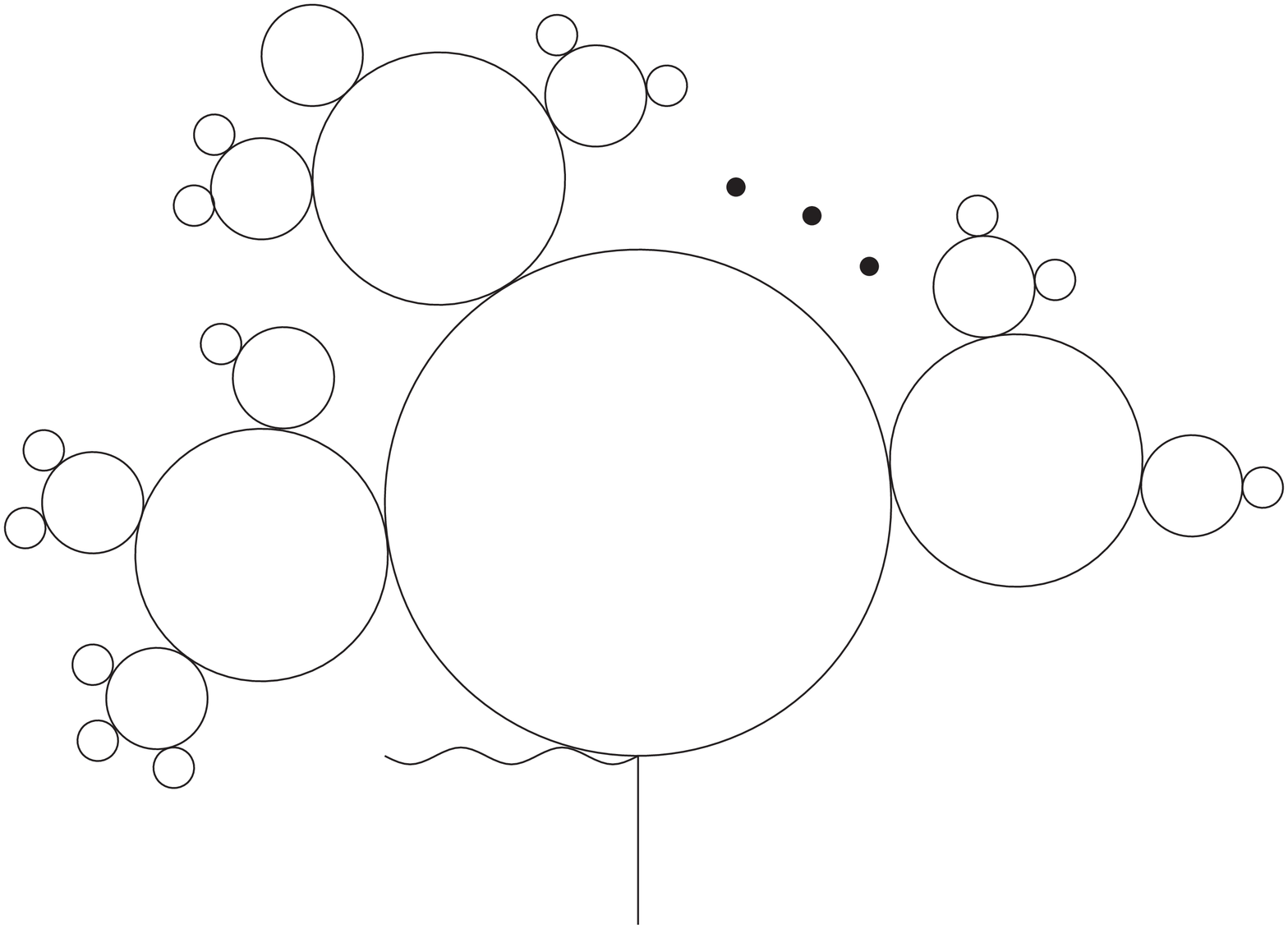}}
$\displaystyle=i$
\raisebox{-0.4cm}{
\includegraphics[keepaspectratio,width=0.05\textwidth,angle=0]{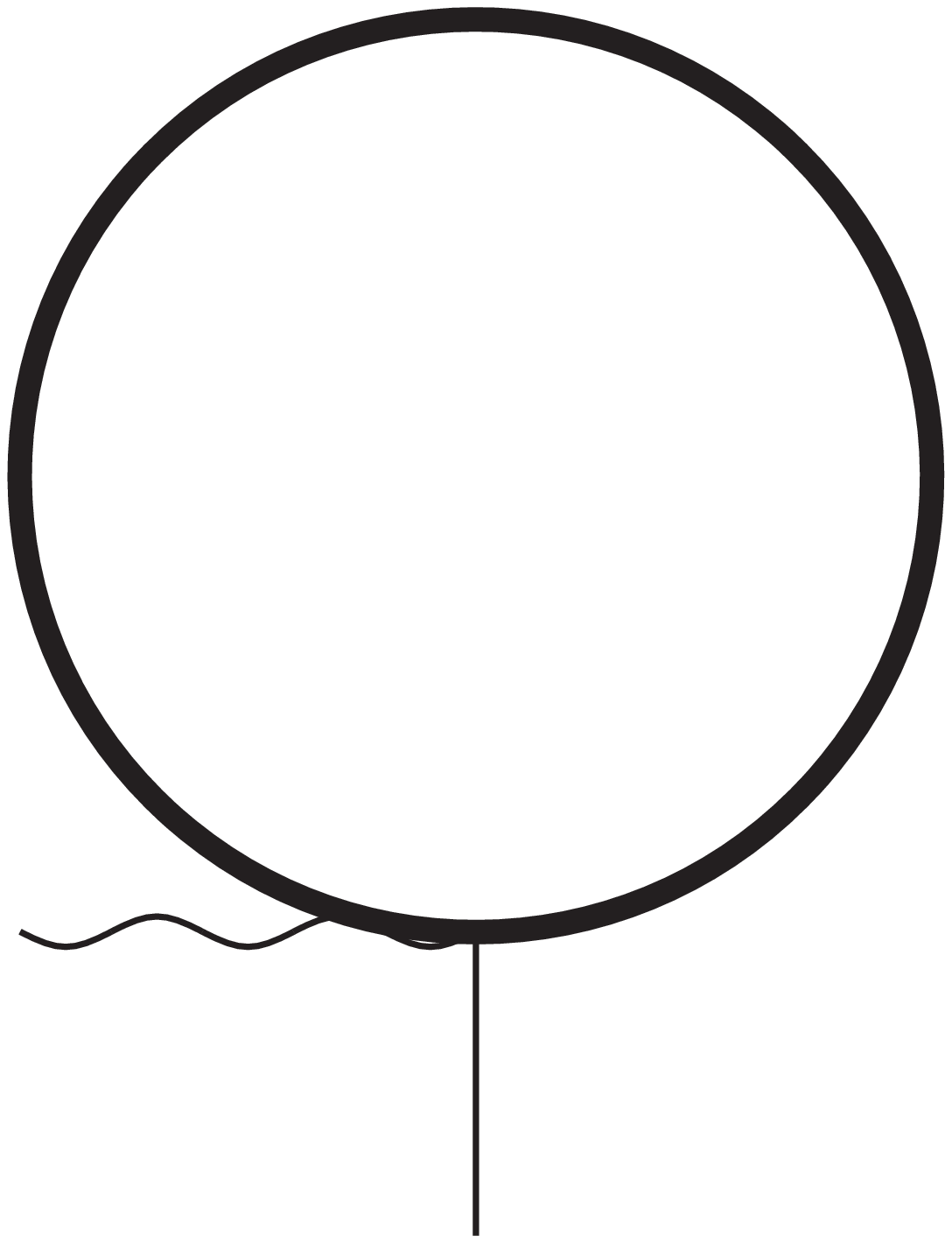}}
\qquad\qquad
$\displaystyle i\sum_{\tn{loops}}$
\raisebox{-0.4cm}{\includegraphics[keepaspectratio,width=0.15\textwidth,angle=0]{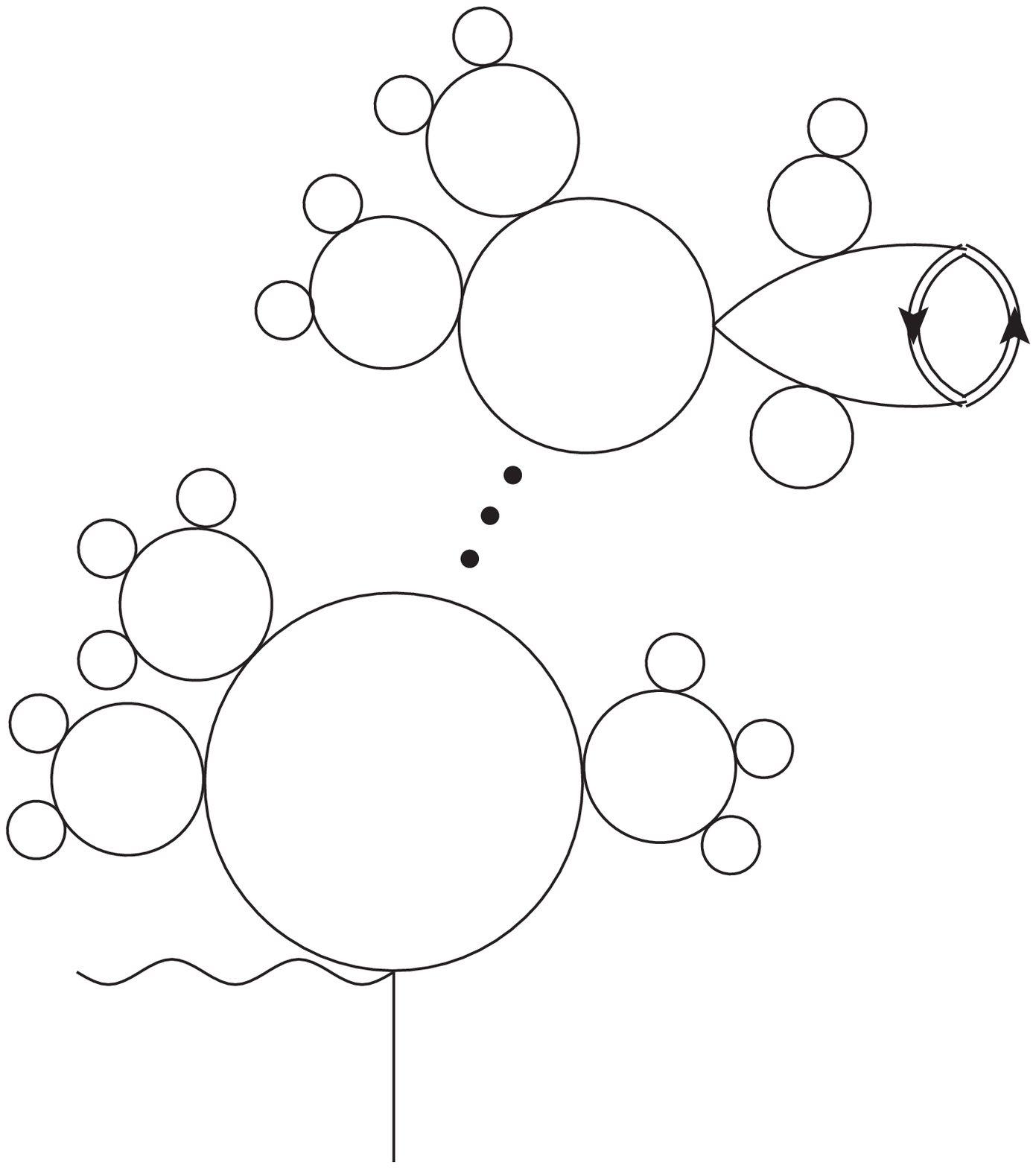}}
$\displaystyle=i\sum_{\tn{skeleton loops}}$
\raisebox{-0.4cm}{\includegraphics[keepaspectratio,width=0.05\textwidth,angle=0]{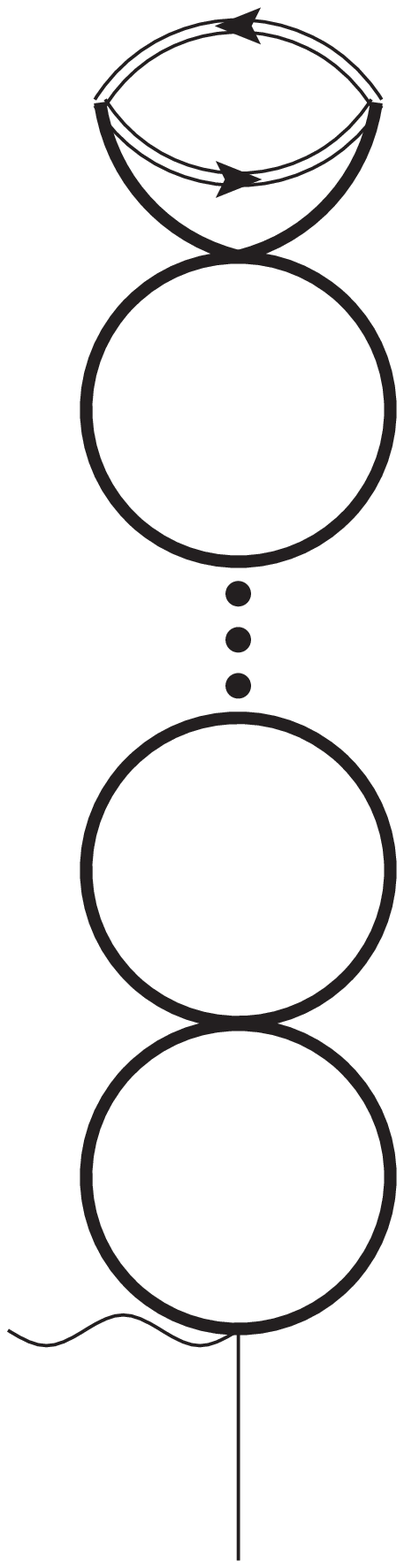}}
\caption{Leading order and next-to-leading order diagrams resummed in the 
equation of state obtained by expanding the self-consistent pion propagator 
to first order in $1/\sqrt{N}.$ The tree-level pion and fermion propagators 
are denoted by thin and double lines, while the thick line represents the 
resummed LO pion propagator.}
\label{Fig:diagrammatic}
\end{figure}

Before closing this section we give in Fig.~\ref{Fig:diagrammatic} the
diagrammatic illustration of the equation of state obtained by a
strict expansion to first order in $1/\sqrt{N}$ of the self-consistent
pion propagator.  This corresponds to the fourth approximation to the
pion propagator discussed in Sec.~\ref{ss:approx}.  Because we
do not draw the counterterm diagrams, we actually obtain the
unrenormalized EoS which reads
\bse
\bea
&&M^2_\tn{LO}+\frac{1}{\sqrt{N}}M^2_\tn{NLO}=\frac{h}{v},
\label{Eq:d1}\\
&&\qquad M^2_\tn{LO}=m^2+\frac{\lambda}{6}\big[v^2+T(M_\tn{LO})\big],
\label{Eq:d2}\\
&&\qquad M^2_\tn{NLO}=\frac{4 g^2 N_c}{1-\frac{\lambda}{6} I(0;M_\tn{LO})}
\left[-\tilde T(m_q)+\frac{\lambda}{12} J(M_\tn{LO},m_q)\right],
\label{Eq:d3}
\eea
\label{Eq:d_all}
\ese
where $I(0;M_\tn{LO})=d T(M)/(d M^2),$ with $T(M)$ defined in 
(\ref{Eq:Tad_pi}). The first set of diagrams are the ${\cal O}(\sqrt{N})$ 
superdaisy diagrams made of pions with tree-level propagators. Their 
resummation is clearly provided by (\ref{Eq:d2}), as one can check iteratively.
The second set of diagrams is ${\cal O}(1)$ and contains only a single
insertion of a fermion bubble. Using Feynman rules, one can readily
check that when the chain of pion bubbles is resummed one obtains
\be
-\frac{\lambda g^2 v/6}{1-\frac{\lambda}{6} I(0;M_\tn{LO})} 
\int_k G_\textnormal{LO}^2(k) \int_p \tr_{c,D}[\gamma_5 D(p)\gamma_5 D(k+p)]\ 
=\frac{4 N_c \lambda g^2 v/6}{1-\frac{\lambda}{6} I(0;M_\tn{LO})} 
\bigg[-I(0;M_\tn{LO}) \tilde T(m_q)+ \frac{1}{2} J(M_\tn{LO},m_q)
\bigg].
\ee
Adding this to the fermion tadpole and dividing by $v$ one obtains the
expression of $M^2_\tn{NLO}$ given in (\ref{Eq:d3}). The finite
version of the EoS (\ref{Eq:d_all}) is obtained by replacing the
integrals by their finite parts taken from the Appendix.

If one would try to use the method described in this section to renormalize 
the EoS (\ref{Eq:EoS}) using the local pion propagator (\ref{Eq:Gp_local}) 
with a mass determined from the gap equation (\ref{Eq:gap_pole}), 
one would encounter a subdivergence proportional to 
$\tilde I_F(M,{\bf 0};m_q)$ which is not canceled. 
This is an artifact of the local approximation used and a self-consistent 
treatment of the propagator would unfold this into renormalizable pieces as 
happened already when the propagator was expanded consistently in $1/\sqrt{N}.$
As we have seen, in this case only subdivergences proportional with $v^2$ and 
$\tilde T_F(m_q)$ appeared. As a consequence, we will not attempt to explicitly 
construct the counterterms when using other approximate forms of the
pion propagator given in Sec.~\ref{ss:approx}. 
Fortunately, in these cases, since the propagators are of a tree-level form, 
the finite part of the integrals can be easily determined and we assume that
in a given equation the subtraction of the infinite part of an integral can 
be achieved by a corresponding subset of the full series of  
counterterm diagrams.

\section{The $\bm{\mu_q-T}$ Phase diagram}

The thermodynamics is determined by solving the field equations, 
{\it i.e.}  the EoS (\ref{Eq:EoS}) and the equations giving the
dependence on $T$ and $\mu_q$ of the two real mean fields $\Phi$ and
$\bar\Phi.$ When the full fermion propagator is replaced by the
tree-level one, one has in view of
Eqs.~(\ref{Eq:I1_def})-(\ref{Eq:I1_final}) and
(\ref{Eq:I2})-(\ref{Eq:T2_decompose})
\bea
\nonumber
&&\frac{d U(\Phi,\bar\Phi)}{d \Phi}
-2 N_c \sqrt{N}  \int \frac{d^3 \k}{(2\pi)^3}
\frac{k^2}{3 E_k} \left(\frac{d \tilde f_\Phi^+(E_k)}{d \Phi}+
\frac{d \tilde f_{\bar\Phi}^-(E_k)}{d \Phi}
\right)\\
&&+g^2\sqrt{N} N_c \left[ 
2 \left(\tilde T_F^0(m_q) - T_F(M)\right)
\frac{d \tilde T^\beta(m_q)}{d \Phi}+
\frac{d \tilde T_2^{\beta,2}(m_q)}{d \Phi}
-M^2\left(\frac{d S^{\beta,1}(M,m_q)}{d \Phi}
+\frac{d S^{\beta,2}(M,m_q)}{d \Phi}\right)
\right]
=0,
\label{Eq:dU_dPhi}
\eea
where $E_k=(\k^2+m_q^2)^{\frac{1}{2}}$ and $M$ satisfies either one of
the gap equations (\ref{Eq:gap_pole}), (\ref{Eq:gap_p0}), or
(\ref{Eq:M2_LO}), or the relation $M^2=h/v.$ The other equation is
similar to (\ref{Eq:dU_dPhi}), the only difference is that the
derivative is taken with respect to $\bar\Phi.$ The integral in
(\ref{Eq:dU_dPhi}) is the contribution of the fermionic trace-log
integral defined in Eq.~(\ref{Eq:I1_def}), while the term proportional
with $g^2$ is the contribution of the quark-pion two-loop integral in
(\ref{Eq:Omega_2PI}) given in Eq.~(\ref{Eq:I2}).  When solving the
field equations for $\Phi$ and $\bar\Phi$, we disregard for simplicity
the contribution of the setting-sun and keep only the one-loop
contribution coming from the fermionic trace-log. The complete
equation (\ref{Eq:dU_dPhi}) is solved only in one case (see the last
row of Table~\ref{tab:phys_data}) in order to estimate the error made
by neglecting this term in all the other cases.  To solve the field
equations we use for the pion propagator a given approximation
described in Sec.~\ref{ss:approx} as will be specified below.

The tricritical point (TCP) and the critical end point (CEP) are
identified as the points along the chiral phase transition line of the
$\mu_q-T$ phase diagram where a first order phase transition turns
with decreasing $\mu_q$ into a second order or crossover transition,
respectively. In case of a crossover, the temperature $T_\chi$ of the
chiral transition is defined as the value where the derivative $d v/d
T$ has a minimum (inflection point of $v(T)$), while the temperature
$T_d$ of the deconfinement transition is obtained as the location of
the maximum in $d\Phi/d T.$ The transition point in the case of a
first order phase transition is estimated by the inflection point
located between the turning points of the multivalued curve $v(\mu_q)$
obtained for a given constant temperature. Although the precise
definition of the 1st order transition point is given by that value of
the intensive parameter for which the two minima of the effective
potential are degenerate, we adopt the definition based on the
inflection point, which is also commonly used in the literature,
because we are not computing the effective potential, but only its
derivatives with respect to the fields and propagators.

\subsection{Phase transition in the chiral limit}

In the chiral limit we solve the EoS (\ref{Eq:EoS}) using only the
local approximation to the pion propagator (\ref{Eq:Gp_local}) with
$M^2=0.$ The critical temperature of the chiral transition $T_\chi$
and the pseudocritical temperature $T_d$ of the deconfinement
transition at vanishing chemical potential, and the location of the
TCP are summarized in Table~\ref{tab:chiral_data} for various forms of
the Polyakov-loop potential. On one hand, one can see that with the
inclusion of the Polyakov loop $T_\chi(\mu_q=0)$ and $T_\tn{TCP}$
increase significantly compared with the values obtained earlier in
\cite{toni04} without the Polyakov loop. On the other hand, in all
cases, the inclusion of the Polyakov loop has little effect on the
value of $\mu_q^\tn{TCP}.$ The increase in $T_\chi(\mu_q=0)$ obtained
with the inclusion of the Polyakov-loop effective potential is
basically determined by the value of its parameter $T_0$, while the
value of $T_\tn{TCP}$ shows no significant variation among different
cases having the same value of $T_0$. One can also see, that as
explained in \cite{fukushima08}, the use of the polynomial and
logarithmic effective potentials for the Polyakov loop, that is
(\ref{Eq:P_eff_pot_poly}) and (\ref{Eq:P_eff_pot_log}), drags the
value of $T_\chi(\mu_q=0)$ closer to the value of the parameter $T_0$
than the use of $U_\tn{Fuku}(\Phi,\bar\Phi)$ given in
(\ref{Eq:P_eff_pot_Fuku}). In this latter case one obtains the
smallest value for $T_\tn{TCP}.$

\begin{table}[!t]
\centering
\begin{tabular}{|c|c||c|c|r|c|}
\hline
$U(\Phi,\bar\Phi)$ & $\ \ T_0\ \ $ & $T_\chi(\mu_q=0)$ & $T_d(\mu_q=0)$ & $\ \ T_\tn{TCP}\ $ & $\ \ \mu_q^\tn{TCP}\ \ $ \\ \hline \hline
$-$           & $-$ & 139.0 &  $-$  &  60.7\ \  & 277.0 \\ \hline \hline
poly          & 270 & 185.6 & 229.0 & 104.5\ \  & 261.8 \\ \hline
poly          & 208 & 168.2 & 176.5 &  96.2\ \  & 263.4 \\ \hline
log           & 270 & 191.4 & 209.0 & 109.4\ \  & 261.2 \\ \hline
log           & 208 & 167.6 & 162.4 & 102.6\ \  & 261.2 \\ \hline
log  & $T_0(\mu_q)$ & 167.9 & 162.8 &  84.3\ \  & 266.9 \\ \hline
Fuku          & $-$ & 176.5 & 193.0 &  99.8\ \  & 262.2 \\ \hline
\end{tabular}
\caption{
The critical temperature $T_\chi$ of the chiral transition and the 
pseudocritical temperature $T_d$ of the deconfinement transition at 
$\mu_q=0,$ and the location of the TCP in units of MeV obtained in 
the chiral limit without the Polyakov loop (first row) and with the 
inclusion of the Polyakov loop using various effective potentials 
summarized in Sec.~\ref{ss:PEP}.
}
\label{tab:chiral_data}
\end{table}

For $T_0=270$~MeV the deconfinement transition line in the $\mu_q-T$
phase diagram is above the chiral transition line in all three
variants of the effective potential for the Polyakov loop. This is
illustrated in the left panel of Fig.~\ref{Fig:chiral_PD} in case of
the polynomial effective potential, where the chiral phase diagram is
compared to the one obtained without including the Polyakov loop.
When the logarithmic effective potential $U_\tn{log}(\Phi,\bar\Phi)$
is used either with a constant $T_0=208$~MeV or with the
$\mu_q$-dependent $T_0$ proposed in \cite{schaefer07} one finds
$T_d<T_\chi$ at $\mu_q=0,$ but at a given value of the chemical
potential the deconfinement transition line crosses the chiral
transition line and remains above it for higher values of $\mu_q.$
This is shown in the right panel of Fig.~\ref{Fig:chiral_PD}, where
the deconfinement transition line is obtained from the inflection
point of $\Phi(T).$ The transition line obtained from the inflection
point of $\bar\Phi(T)$ is practically indistinguishable from the line
shown in the figure.  One can see that in contrast to the case of
constant $T_0,$ where basically the deconfinement transition line is
not affected by the increase of $\mu_q,$ with a $\mu_q$-dependent
$T_0$ the deconfinement transition line strongly bends, staying close
to the chiral line. The two lines cross just above the TCP.

The lowering of the deconfinement transition in the case when
$T_0(\mu_q)$ is used and as a result the shrinking of the so-called
quarkyonic phase was already observed in Ref.~\cite{abuki08}. As
distinguished from the mesonic phase which is confined and has zero
quark number density and the deconfined phase which has finite quark
number density, the quarkyonic phase is a confining state made of
quarks and is characterized by a high quark number density and
baryonic (three-quark state) thermal excitations. Based on the fact
that in the PNJL model the quantity measuring the quark content inside
thermally excited particles carrying baryon number shows a pronounced
change along the chiral phase transition line, the region of the
$\mu_q-T$ plane for which $T_\chi<T<T_d$ was identified in
\cite{fukushima08} with the quarkyonic phase. The first numerical evidence
from lattice QCD for the existence of a phase which is neither the
hadronic nor the deconfined phase and is characterized by a high value
of the quark number density was given in \cite{fodor07}. This could be
a candidate for the quarkyonic phase. Further evidence for such a
state was reported also in \cite{miura09} within the strong-coupling
expansion of the lattice QCD.

\begin{figure}[!t]
\begin{center}
\includegraphics[keepaspectratio,width=0.495\textwidth,angle=0]{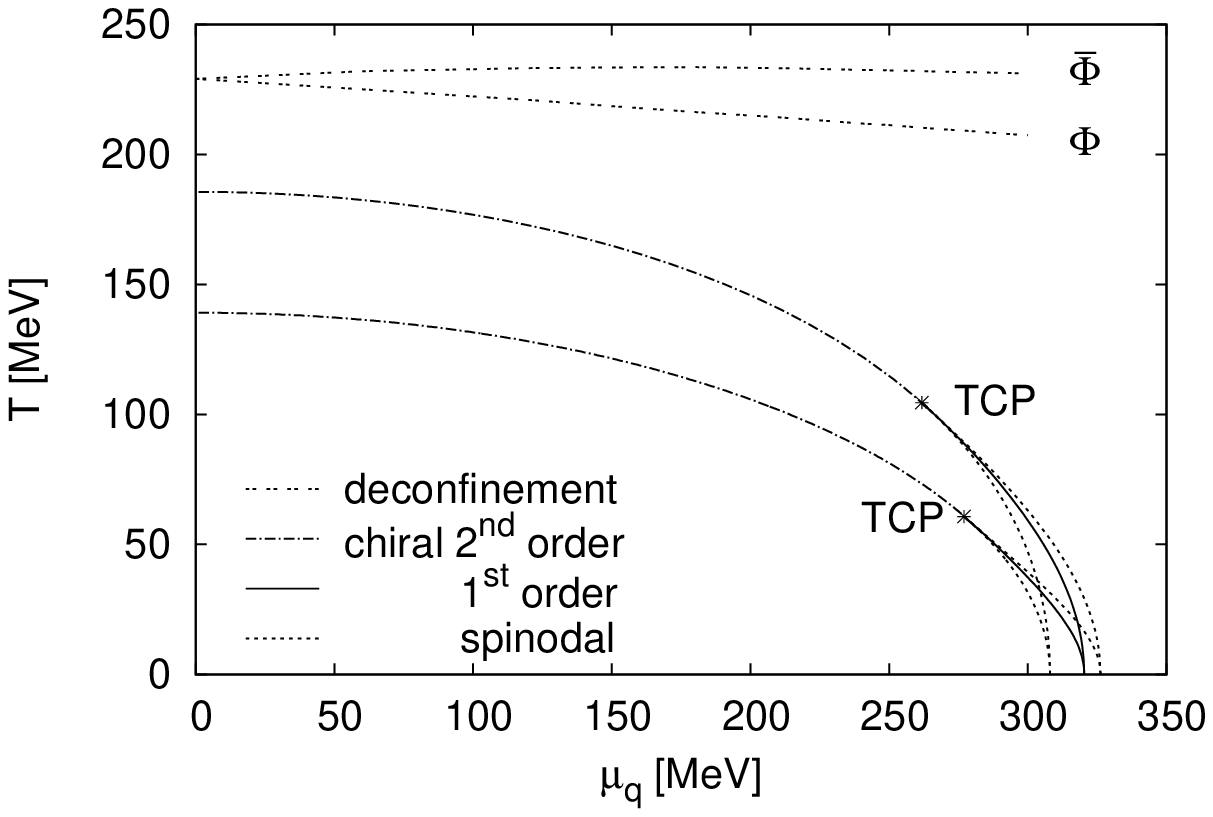}
\includegraphics[keepaspectratio,width=0.495\textwidth,angle=0]{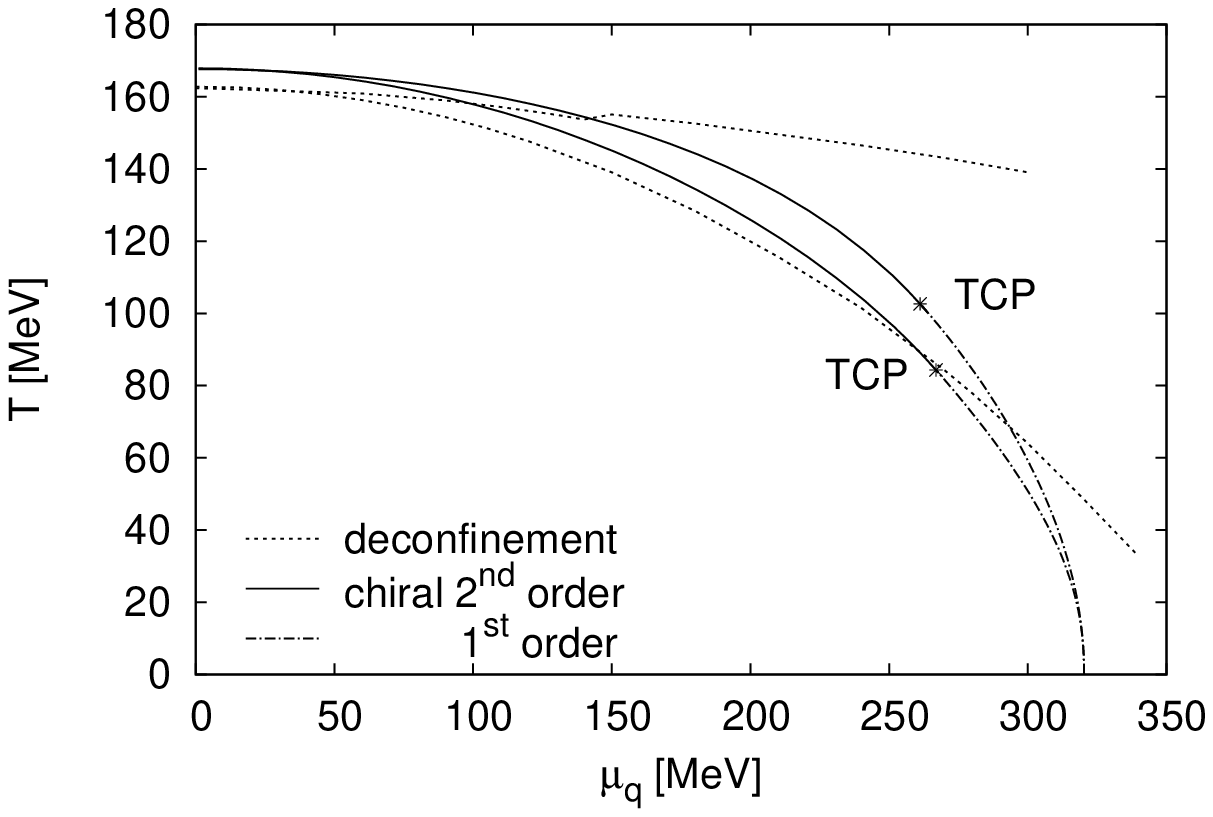}
\caption{Left panel: Phase diagrams obtained in the chiral limit without and 
with the inclusion of the Polyakov loop. The latter has higher $T_\tn{TCP}$ 
and was obtained using $U_\tn{poly}(\Phi,\bar\Phi)$ with $T_0=270$~MeV.
Shown are the location of the inflection points of $\Phi(T)$ and 
$\bar\Phi(T).$ Right panel: Chiral and deconfinement phase 
transitions obtained for $U_\tn{log}(\Phi,\bar\Phi)$ with $T_0=208$~MeV 
(upper curves) and with $T_0(\mu_q)$ (lower curves). The deconfinement 
transition line is obtained from the inflection point of $\Phi(T).$
}
\label{Fig:chiral_PD}
\end{center}
\end{figure}

Comparing our results on the phase diagram to those obtained in the
chiral limit of the PNJL model one can notice differences of both
qualitative and quantitative nature. In the nonlocal PNJL model of
Ref.~\cite{sasaki07} the deconfinement phase transition line starts at
$\mu_q=0$ below the chiral transition line both for a polynomial and a 
logarithmic Polyakov-loop effective potential with $T_0=270$~MeV, so
that the two transition lines cross at finite $\mu_q.$ In our case
this happens only for the logarithmic potential with $T_0=208$~MeV, as
can be seen in Fig.~\ref{Fig:chiral_PD}. In \cite{sasaki07,costa09a}
the values of $T_\chi(\mu_q=0)$ and $T_\tn{TCP}$ are much larger than
in our case, while the value of $\mu_q^\tn{TCP}$ is similar to ours.

\subsection{Phase transition in case of the physical pion mass \label{ss:phys}}

In the case of the physical pion mass we solve the EoS (\ref{Eq:EoS})
using each one of the four approximations to the pion propagator
introduced in Sec.~\ref{ss:approx}. Within the approximation
corresponding to a strict expansion in $1/\sqrt{N}$ of the pion
propagator and of the EoS discussed in detail in
Sec.~\ref{sec:renorm} there is no CEP in the $\mu_q-T$ phase diagram
within a range $0<\mu_q<500$~MeV.  Without inclusion of the Polyakov
loop the transition at $\mu_q=0$ is a very weak crossover
characterized by a large value of the width at half maximum of
$-dv/dT,$ $\Gamma_\chi\sim 100$~MeV.  Including the Polyakov loop,
although the width at half maximum of $-dv/d T$ decreases by a factor
of 2 compared to the case without the Polyakov loop, the transition
remains a crossover for $\mu_q<500$~MeV.  This means that as a result
of resumming in the pion propagator ${\cal O}(1/\sqrt{N})$ fermionic
fluctuations, while keeping the fermion propagator unresummed, the
crossover transition at $\mu_q=0$ softens and increasing $\mu_q$
cannot turn the phase transition into a first order one. For the other
three approximations, which all resum infinitely many orders in
$1/\sqrt{N},$ the phase transition turns with increasing $\mu_q$ from
a crossover type into a first order transition and in consequence
there is a CEP in the $\mu_q-T$ phase diagram.  For these cases the
result is summarized in Table~\ref{tab:phys_data} for various forms of
the Polyakov-loop potential reviewed in Sec.~\ref{ss:PEP}.

\begin{table}[htbp]
\centering
\begin{tabular}{|c|c|c||c|c|c|c|c|}
\hline
$U(\Phi,\bar\Phi)$ & $\ \ T_0\ \ $ & $\ \ \ G_\pi(p)\ \ \ $ & $T_\chi(\mu_q=0)$ & $T_d(\mu_q=0)$ & $\Gamma_\chi$ & $\ \ T_\tn{CEP}\ \ $ & $\ \ \mu_q^\tn{CEP}\ \ $ \\ \hline \hline
$-$  & $-$ & local, pole  & 152.8 &  $-$  & 37.6 & 14.4 & 327.1 \\ \hline
$-$  & $-$ & local, $p=0$ & 158.2 &  $-$  & 41.5 & 12.1 & 329.1 \\ \hline
$-$  & $-$ & large-$N$    & 158.6 &  $-$  & 40.7 & 13.5 & 328.6 \\\hline\hline
poly & 270 & local, pole  & 205.6 & 226.8 & 25.6 & 37.8 & 326.9 \\ \hline
poly & 208 & local, pole  & 180.6 & 175.0 & 19.8 & 35.3 & 326.7 \\ \hline
poly & 270 & local, $p=0$ & 211.4 & 217.8 & 27.3 & 32.4 & 329.0 \\ \hline
poly & 208 & local, $p=0$ & 184.6 & 176.7 & 22.7 & 30.1 & 328.9 \\ \hline
poly & 270 & large-$N$    & 212.5 & 217.4 & 28.3 & 32.9 & 328.8 \\ \hline
poly & 208 & large-$N$    & 184.6 & 176.8 & 22.3 & 30.6 & 328.8 \\ \hline
log  & 270 & local, pole  & 207.2 & 207.7 & 12.3 & 39.3 & 327.0 \\ \hline
log  & 208 & local, pole  & 168.0 & 167.0 & *30.3& 37.9 & 326.9 \\ \hline
log  & 270 & local, $p=0$ & 209.8 & 209.3 & 12.1 & 33.9 & 329.1 \\ \hline
log  & 208 & local, $p=0$ & 168.5 & 167.0 & *42.8& 32.7 & 329.0 \\ \hline
log&$T_0(\mu_q)$&local, $p=0$&168.9& 167.4 & *42.5& 25.7 & 328.7 \\ \hline
log  & 270 & large-$N$    & 209.7 & 209.3 & 12.0 & 34.5 & 329.0 \\ \hline
log  & 208 & large-$N$    & 168.5 & 167.1 & *43.0& 33.0 & 328.9 \\ \hline
Fuku & $-$ & local, pole  & 191.0 & 188.7 & 19.8 & 36.2 & 326.8 \\ \hline
Fuku & $-$ & local, $p=0$ & 195.3 & 191.2 & 21.2 & 31.2 & 328.9 \\ \hline
Fuku & $-$ & large-$N$    & 195.2 & 191.3 & 21.2 & 31.8 & 328.8 \\ \hline \hline
poly & 208 & large-$N$, full & 188.1 & 183.1 & 21.4 & 32.2 & 329.0 \\ \hline
\end{tabular}
\caption{The pseudocritical temperatures $T_\chi$ and  $T_d$ of the chiral 
and deconfinement transitions, the half-width at half maximum 
$\Gamma_\chi$ of $-d v/d T$ at $\mu_q=0$ 
(in the cases marked with $*$, due to an asymmetric shape of $-d v/d T$ the 
full width is given) 
and the location of the CEP in units of MeV 
obtained in various approximations for the pion propagator without and 
with the inclusion of the Polyakov loop. The two local approximations are 
defined by (\ref{Eq:gap_pole}) and (\ref{Eq:gap_p0}), respectively. 
The large-$N$ approximation is  defined by (\ref{Eq:Gp_third}) and 
(\ref{Eq:EoS_third}). Only for the result in the last row the contribution 
of the setting-sun was kept in (\ref{Eq:dU_dPhi}).}
\label{tab:phys_data}
\end{table}

In the cases studied in Table~\ref{tab:phys_data} increasing $\mu_q$
drives at $T=0$ the restoration of chiral symmetry via a first order
transition at some value $\mu_q^c>M_q.$ Increasing the temperature
$\mu_q^c$ decreases and the first order chiral restoration becomes a
crossover at a much lower temperature $T_\tn{CEP}$ than in the chiral
case. The inclusion of the Polyakov loop increases significantly the
value of $T_\tn{CEP},$ but as in the chiral case it has little effect
on the value of $\mu_q^\tn{CEP}.$ One can see that neither the choice
of the effective potential for the Polyakov loop nor the value of
$T_0$ has a significant effect on the value of $\mu_q^\tn{CEP}.$ Some
variation can be observed among the values of $\mu_q^\tn{CEP}$
obtained using different approximations for the pion propagator. The
result in the last row was obtained by keeping in the field equation
of the Polyakov loop (\ref{Eq:dU_dPhi}) and its conjugate the
contribution of the setting-sun diagram, while in all other cases only
the contribution of the fermionic trace-log was kept. Comparing the
result in the last row with that of the last row obtained using the
polynomial Polyakov-loop potential, one can see that the error we make
by neglecting the setting-sun contribution in all other cases is
fairly small.

The values of $T_\chi$ and $T_d$ at $\mu_q=0$ are mostly influenced by
the choice of the Polyakov effective potential and the value of $T_0:$
they decrease with the decrease of $T_0$ and by using the logarithmic
potential instead of the polynomial one. Using the polynomial
potential with $T_0=270$~MeV the confinement transition line in the
$\mu_q-T$ plane is above the chiral transition line. This can be seen
in the left panel of Fig.~\ref{Fig:phys_PD} where the phase diagram is
compared with the one obtained without the inclusion of the Polyakov
loop. As in the chiral case, when a logarithmic potential is used with
either a fixed value $T_0=208$~MeV or with a $\mu_q$-dependent $T_0,$
the deconfinement transition line starts at $\mu_q=0$ below the chiral
one and the two lines cross at some higher value of $\mu_q.$ This can
be seen in the right panel of Fig.~\ref{Fig:phys_PD}. When
$T_0(\mu_q)$ is used the two lines go together until they cross each
other just above the location of the CEP. This $\mu_q$-dependent $T_0$
gives the lowest value of $T_\tn{CEP},$ similar to the results
reported in \cite{ciminale08} and \cite{herbst10}. Because of the much
lower value of the $T_\tn{CEP}$ the shrinking of the quarkyonic phase
is more pronounced than in the chiral case, as the deconfinement
transition lines approaches the $\mu_q$ axis.  This is even more the
case here, with a physical pion mass, since the deconfinement
transition is a crossover and as such it happens in a relatively large
temperature interval. However, the quarkyonic phase does not vanish
completely as happens in \cite{herbst10}, where quantum fluctuations
are included using functional renormalization group methods.

\begin{figure}[!t]
\centering
\includegraphics[keepaspectratio,width=0.495\textwidth,angle=0]{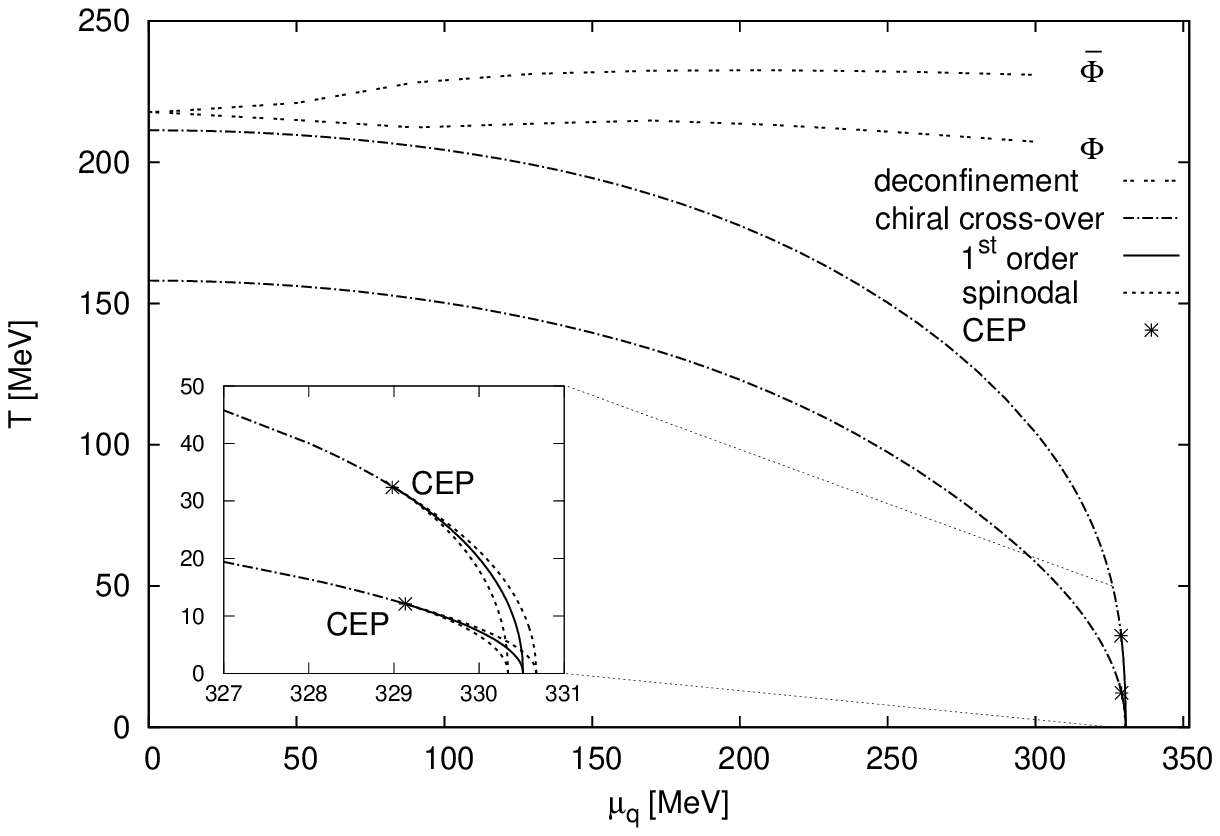}
\includegraphics[keepaspectratio,width=0.495\textwidth,angle=0]{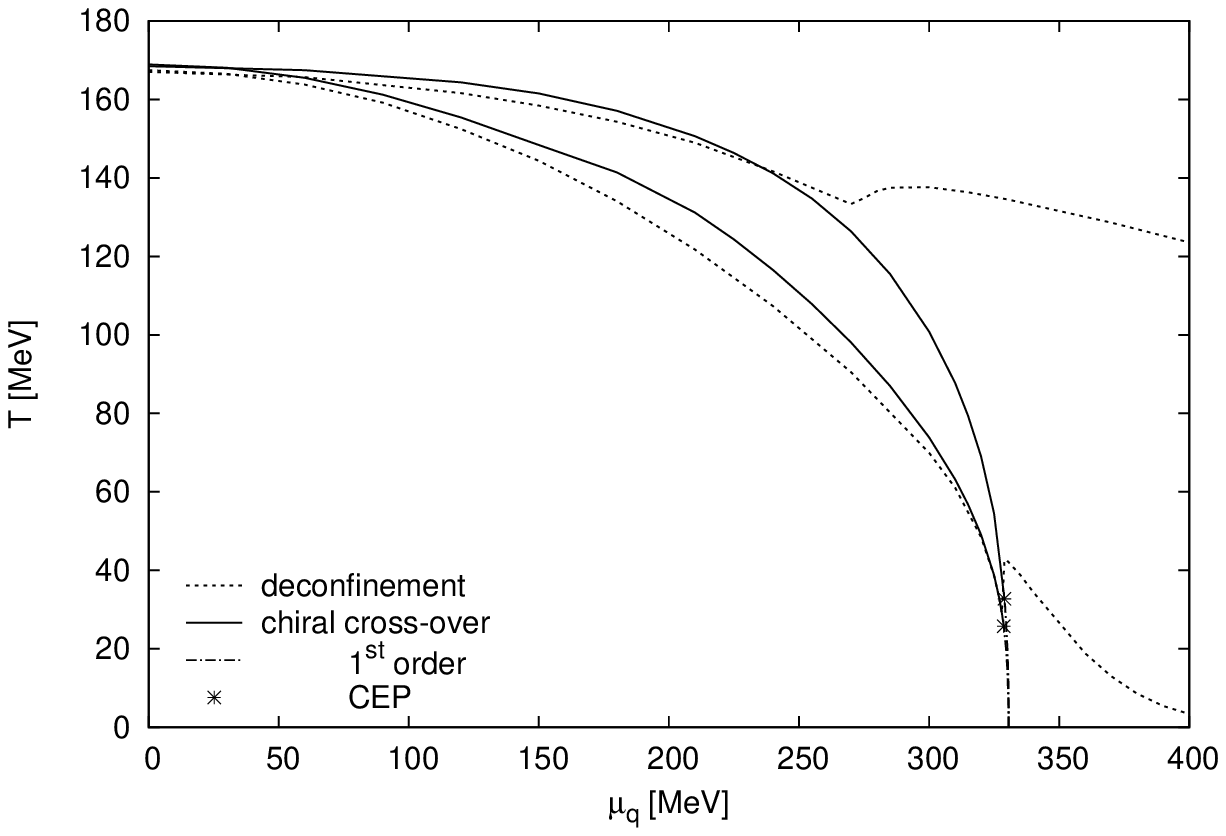}
\caption{Left panel: Phase diagrams obtained for the 
physical value of the pion mass using the local approximation to $G_\pi$ with 
a mass determined by (\ref{Eq:gap_p0}) without and with the inclusion of 
the Polyakov loop. The latter has higher $T_\tn{CEP}$ and was obtained using 
$U_\tn{poly}(\Phi,\bar\Phi)$ with $T_0=270$~MeV. The part of the phase 
diagram where the transition is of first order is enlarged in the 
inset. Shown are the global maxima of $d\Phi(T)/d T$ and 
$d\bar\Phi(T)/d T.$ Right panel: Chiral and deconfinement phase 
transitions obtained for $U_\tn{log}(\Phi,\bar\Phi)$ with $T_0=208$~MeV 
(upper curves) and with $T_0(\mu_q)$ (lower curves). The deconfinement 
transition line is obtained from the global maximum of $d\Phi(T)/d T.$
}
\label{Fig:phys_PD} 
\end{figure}

\begin{figure}[!b]
\centering
\includegraphics[keepaspectratio,width=0.495\textwidth,angle=0]{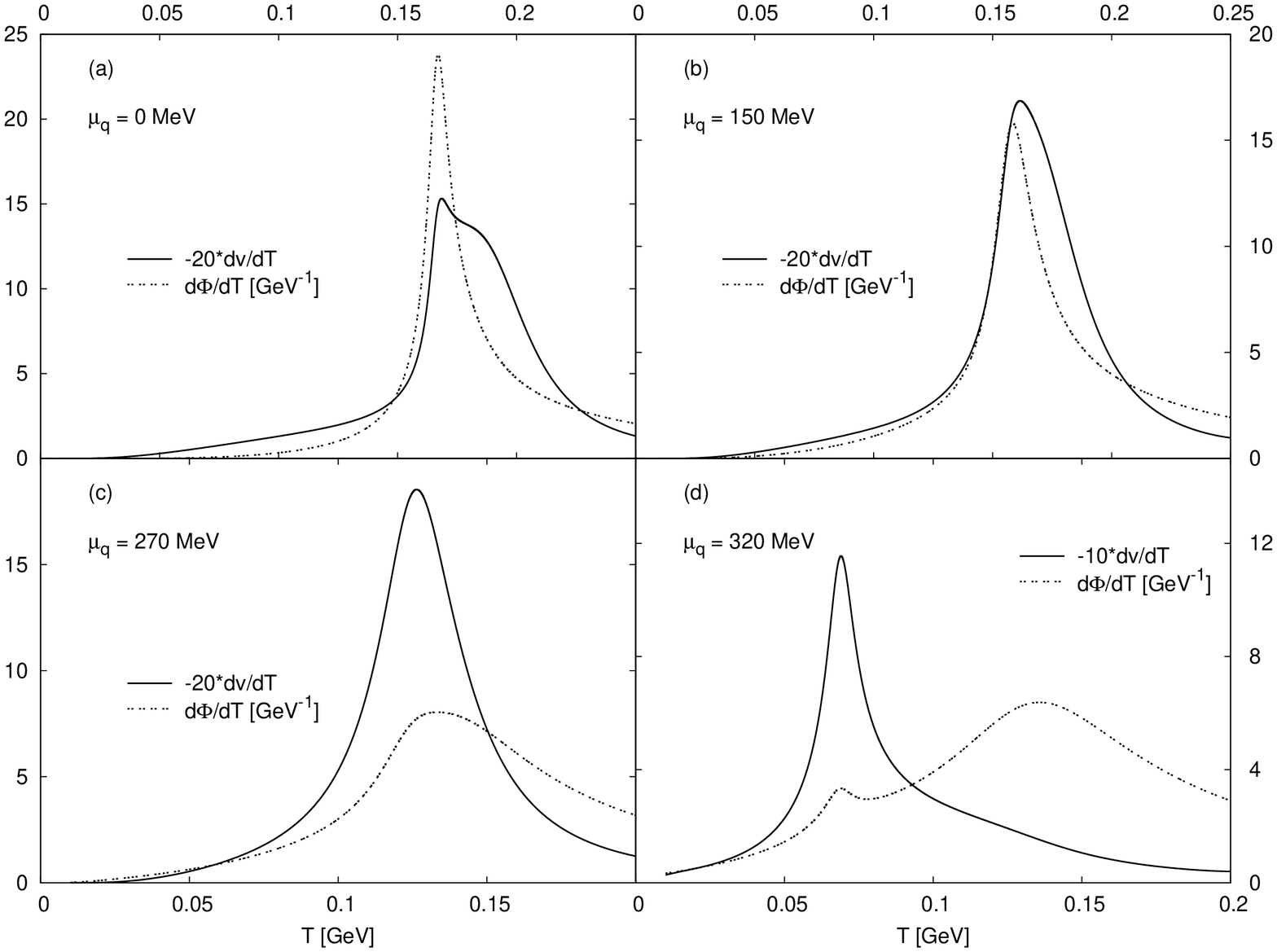}
\includegraphics[keepaspectratio,width=0.495\textwidth,angle=0]{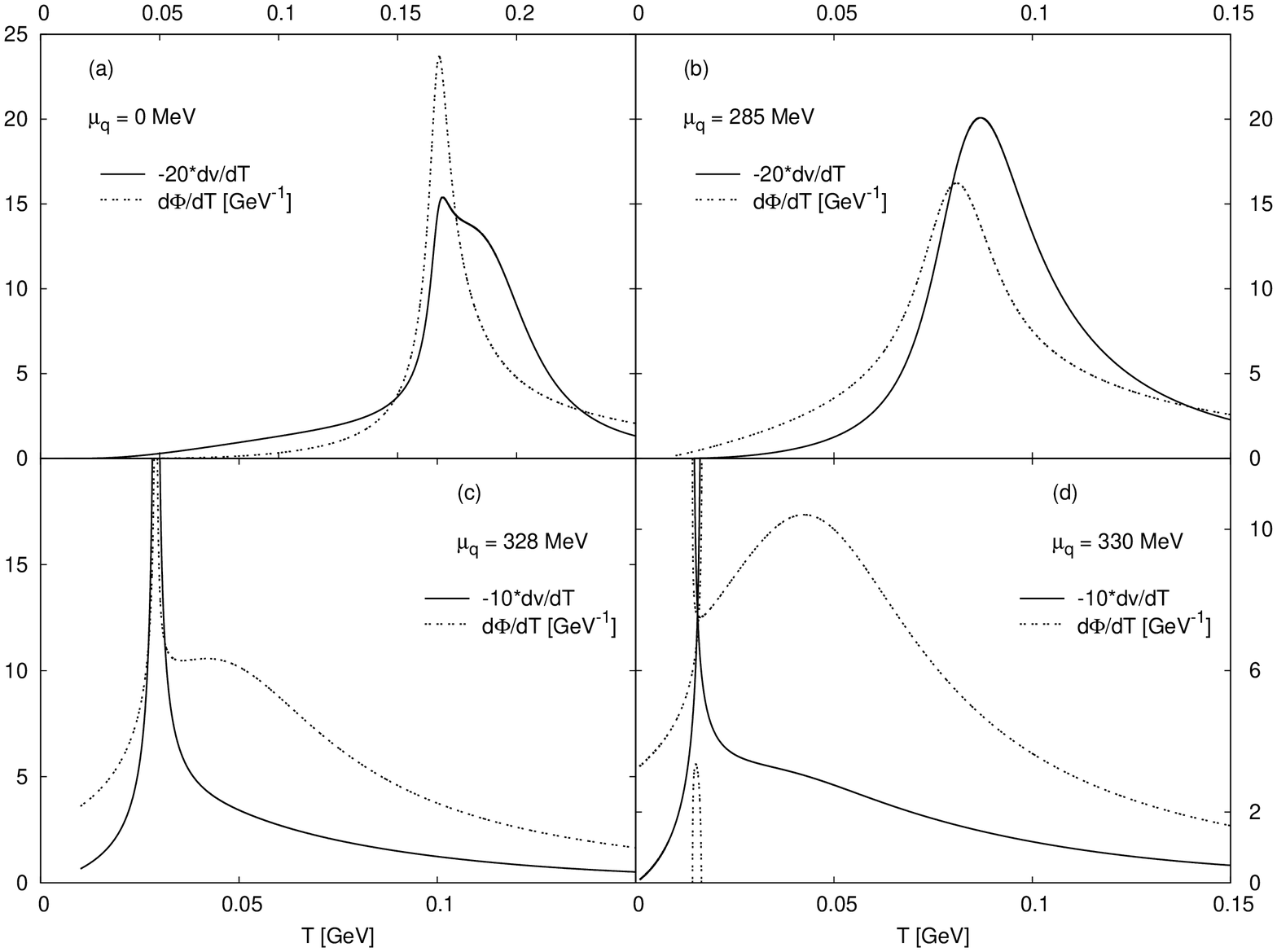}
\caption{Evolution of the maxima of the $-d v/d T$ and $d \Phi/d T$ 
with the increase of the chemical potential $\mu_q$ in case of using
$U_\tn{log}(\Phi,\bar\Phi)$ with a $T_0=208$~MeV (left figure)
and with a $\mu_q$-dependent $T_0$ parameter (right figure).
}
\label{Fig:maxima} 
\end{figure}

By studying the derivatives of the $v(T)$ and $\Phi(T)$ curves one
observes in panel (a) of Fig.~\ref{Fig:maxima} that at low $\mu_q$ it
is the Polyakov loop which plays the driving role in the transition:
for $\mu_q=0$ the $d v/d T$ is much wider and has a small peak in the
temperature range where $\Phi(T)$ shows a pronounced variation.  This
happens only for very low values of $\mu_q,$ as in the region of
$\mu_q$ where the deconfinement transition line is a little bit
further below the chiral transition line than for $\mu_q=0$, such a
driving role cannot be identified. For values of $\mu_q$ where the two
transition line cross and also in the region where the chiral
transition line is below the deconfinement one can see the influence
of the chiral transition on the shape of $d\Phi/d T.$ This is the most
pronounced in the case of the $\mu_q$-dependent $T_0$ where the two
transition lines cross near the CEP. In this case the chiral phase
transition plays the driving role as one can clearly see on panel (c)
of Fig.~\ref{Fig:maxima} (right). In panel (d) of
Fig.~\ref{Fig:maxima} (left) one sees that $d\Phi/d T$ has two
peaks. In such cases, as in Ref.~\cite{tuominen08}, the position of
the higher peak is followed to determine the deconfinement transition
temperature, since the first peak is a result of the influence of the
chiral phase transition.

From Table~\ref{tab:phys_data} one can see that there is a correlation 
between the strength of the chiral crossover at $\mu_q=0$ as measured 
by $\Gamma_\chi$ and the location of CEP: weaker crossover (larger value of
$\Gamma_\chi$) corresponds in general to a larger value of  $\mu_q^\tn{CEP}.$
In the cases marked with an asterisk in Table~\ref{tab:phys_data} a 
$d v/d T$ is distorted by the temperature dependence of the Polyakov loop,
as one can see in panel (a) of Fig.~\ref{Fig:maxima}. For this reason 
in this cases we denote by $\Gamma_\chi$ the full width at half maximum of 
$-d v/d T.$ In other cases one gives the half-width at half maximum. 
This is measured on the left of the maximum, because when the gap equation 
is used, the threshold of the fermionic bubble is generally on the right 
of the maximum and distorts the $d v/d T$ curve.

\section{Discussion and conclusions}

Using the tree-level fermion propagator and several approximate forms
of the pion propagator obtained within a large-$N_f$ expansion, we
studied in the chiral limit and for the physical value of the pion
mass the influence of the Polyakov loop on the chiral phase
transition.  We obtained that only when the local part of the
approximate pion propagator resums infinitely many orders in $1/N_f$
of fermionic contributions it is possible to find a CEP on the
chiral phase transition line of the $\mu_q-T$ phase diagram. When the
logarithmic form $U_\tn{log}(\Phi,\bar\Phi)$ of the effective
potential for the Polyakov loop was used with parameter $T_0=208$~MeV
a crossing between the chiral and deconfinement transition lines was
observed, with the latter line starting at $\mu_q=0$ slightly below
the former one. In this case the existence of the quarkyonic phase is
possible.

We have seen at the beginning of Sec.~\ref{ss:phys} that as a
result of resumming in the pion propagator ${\cal O}(1/\sqrt{N})$
fermionic fluctuations obtained with a strict expansion in
$1/\sqrt{N}$, while keeping the fermion propagator unresummed, the 
phase transition softens. One can easily demonstrate the same feature
by including the contributions of the fermion vacuum fluctuations and
of the pion tadpole in the equation of state of
Ref.~\cite{schaefer07}, that is the field equation determining the
chiral order parameter.  There, because the parameters of the PQM
model were determined at tree-level, the fermionic vacuum fluctuations
coming from the fermion tadpole ($\tilde T^0$) were neglected, while
the pions were treated at tree-level.  However, by choosing an
appropriate renormalization scale one can arrange for the vanishing of
the entire $\tilde T^0_F$ only at $T=0.$ At finite temperature the
vacuum fluctuation is in this way correctly included and due to the
temperature dependent fermionic mass $\tilde T^0_F$ will be
nonvanishing. The value of the renormalization scales for which the
fermion and pion tadpoles vanish at $T=\mu_q=0$ are 
$M_{0F}=\sqrt{e} m_q$ and $M_{0B}=\sqrt{e} m_\pi,$ respectively. The
importance of including the vacuum fluctuations was discussed also in
\cite{nakano10,skokov10b}, where the effect on the location of the CEP
and on the isentropic trajectories in the $\mu_q-T$ plane was shown.
From Table~\ref{tab:compare} one can see that comparing with the
original result of Ref.~\cite{schaefer07} the inclusion of the
fermionic vacuum fluctuations softens the transition at $\mu_q=0$, as
shown by the larger full width $\Gamma_\chi$ at half maximum of 
$-d v/d T,$ and in consequence the location of the CEP is moved to
higher values of $\mu_q$ and lower values of $T.$ Inclusion of the
pion vacuum and thermal fluctuations in the equation of state through
a pion tadpole further accentuates this behavior. Inclusion of the
fluctuations using functional renormalization group methods also
pushed the location of the CEP to higher values of $\mu_q,$ as can be
seen by comparing the left panel of Fig.~6 in ~\cite{herbst10} to
Fig.~6 of \cite{schaefer07}.

\begin{table}[htbp]
\centering
\begin{tabular}{|l|cccc|c|r|c|c|}
\hline
& $\ \tilde T^0_F\ $  & $\ T^0_F\ $ & $\ \tilde T^{\beta}\ $ & $\ T^{\beta}\ $ & $\ T_\chi(\mu_q=0)\ $ & $\ \ \Gamma_\chi\ \ $ & \ \ $T_\tn{CEP}$\ \  & \ \ $\mu_q^\tn{CEP}$\ \ \\
\hline \hline
QP  & $-$ & $-$ & $+$ & $-$ & 184.6 & 4.6  & 162.8 & 165.1 \\ \hline
QP  & $-$ & $-$ & $+$ & $+$ & 180.2 & 8.6  & 145.3 & 204.3 \\
QFT & $+$ & $-$ & $+$ & $-$ & 173.0 & 26.9 & 91.3  & 241.1 \\
QFT & $+$ & $+$ & $+$ & $+$ & 170.1 & 30.3 & 85.5  & 243.5 \\
\hline
\end{tabular}
\caption{
The pseudocritical temperatures of the chiral transition and the full width 
$\Gamma_\chi$ at half maximum of $-d v/ (d T)$ at $\mu_q=0,$ and the 
location of the CEP in units of MeV in various treatments of the model 
with a physical pion mass. The Polyakov loop is included using 
$U_\tn{poly}(\Phi,\bar\Phi)$ 
and $T_0=208$~MeV. QP stands for the quasiparticle approximation 
in which the vacuum fluctuations in the fermion ($\tilde T^0_F$) 
or pion ($T^0_F$) tadpoles are disregarded (marked by $-$) and only 
the finite temperature part of the tadpoles ($\tilde T^{\beta}$ or 
$T^{\beta}$) is kept (marked by $+$). QFT stands for a quantum field 
theoretical calculation where the vacuum fluctuations are properly treated. 
The first row is the reproduced result of \cite{schaefer07}.
}
\label{tab:compare}
\end{table}

It remains to be seen to what extent our results are stable against
the use of the self-consistent propagator for fermions, as required by
a completely systematic large-$N_f$ expansion. A highly interesting
question which requires going beyond the level of approximations of
this work is whether a completely systematic expansion in
$1/\sqrt{N}$ of the propagator equations could lead to the existence
of the CEP in the phase diagram, and how the results obtained within
such a resummation scheme are related to a numerically even more
demanding resummation represented by the complete self-consistent
solution (without further expansion in $1/\sqrt{N}$) 
of the coupled pion and fermion propagator equations.

\begin{acknowledgments} 
The authors benefited from discussions with Andr\'as Patk{\'o}s and 
Antal Jakov\'ac. This work is supported by the Hungarian Research Fund 
under Contracts No.~T068108 and No.~K77534. 
\end{acknowledgments}
   
\appendix*
\section{Integrals and performing the color trace\label{app:integrals}}

\subsection{The trace-log}
We calculate first an integral appearing in (\ref{Eq:Omega_grand_pot}) 
when the approximation $G(k)\to D(k)$ is used. This is defined as
\bea
I_1=\sqrt{N} i\, \tr_{D,c} \int_k \ln D^{-1}(k)=
-2 \sqrt{N} T \sum_{i=1}^{N_c} \sum_n \int \frac{d^3 \k}{(2\pi)^3} 
\ln \left[\beta^2\left(\omega_n^2+E_k^2\right)\right],
\label{Eq:I1_def}
\eea
where we used (\ref{Eq:sum_int_def}) and the notation $E_k^2=\k^2+m_q^2.$
The Matsubara frequencies are $\omega_n=(2 n+1)\pi T-i \mu_i,$ with the 
color-dependent chemical potentials defined in (\ref{Eq:c-dep_mu}).
Doing the Matsubara sum and an integration by parts one obtains
\bea
I_1=-2 \sqrt{N}  \sum_{i=1}^{N_c}\int \frac{d^3 \k}{(2\pi)^3} 
\left[E_k+\frac{k^2}{3 E_k} \left(\tilde f^+_i(E_k)+\tilde f^-_i(E_k)\right)
\right],
\eea
where $\tilde f^\pm_i(E)=1/(\exp[\beta(E\mp \mu_i)]+1).$
Using the diagonal form of the  Polyakov-loop operator given in 
(\ref{Eq:Polyakov_op_diag}) one can define following Ref.~\cite{hansen07}
\bse
\bea
\tilde f_\Phi^+(E)&=&
\frac{1}{N_c}\sum_{i=1}^{N_c}\tilde f_i^+(E)
=\frac{1}{N_c}\tr_c \frac{1}{L e^{\beta(E-\mu_q)}+1},
\label{Eq:F_P+def}
\\
\tilde f_\Phi^-(E)&=&
\frac{1}{N_c}\sum_{i=1}^{N_c}\tilde f_i^-(E)
=
\frac{1}{N_c}\tr_c \frac{1}{L^\dagger e^{\beta(E+\mu_q)}+1}.
\label{Eq:F_P-def}
\eea
\label{Eq:F_P+_P-def}
\ese
Then, simple algebra shows that upon working out the traces
$\tilde f_\Phi^\pm(E)$ can be expressed 
in terms of $\Phi=(\tr_c L)/N_c$ and $\bar\Phi=(\tr_c L^\dagger)/N_c$ as
\bse
\bea
\tilde f_\Phi^+(E)&=&
\frac{\left(\bar\Phi +2\Phi e^{-\beta(E-\mu_q)}\right) e^{-\beta(E-\mu_q)} 
      + e^{-3\beta(E-\mu_q)} }
{1 + 3\left( \bar\Phi + \Phi e^{-\beta(E-\mu_q)} \right) e^{-\beta(E-\mu_q)} 
      + e^{-3\beta(E-\mu_q)}},
\label{Eq:F_P+}\\
\tilde f_{\Phi}^-(E)&=&
\frac{ \left( \Phi + 2\bar\Phi e^{-\beta(E+\mu_q)} \right) e^{-\beta(E+\mu_q)} 
      + e^{-3\beta(E+\mu_q)} }
{1 + 3\left( \Phi + \bar\Phi e^{-\beta(E+\mu_q)} \right) e^{-\beta(E+\mu_q)} 
      + e^{-3\beta(E+\mu_q)}} .
\label{Eq:F_P-}
\eea
\label{Eq:F_P+_P-}
\ese
Through these functions the integral $I_1$ in (\ref{Eq:I1_def}) is also 
expressed in terms of $\Phi$ and $\bar\Phi$ as
\be
I_1=-2 N_c\sqrt{N}  \int \frac{d^3 \k}{(2\pi)^3}
\left[
E_k+\frac{k^2}{3 E_k} \left(\tilde f_\Phi^+(E_k)+
\tilde f_{\bar\Phi}^-(E_k)\right)
\right].
\label{Eq:I1_final}
\ee
The derivative of this integral with respect to $\Phi$ is used in 
(\ref{Eq:dU_dPhi}).

\subsection{Tadpole and bubble integrals}

The pion tadpole integral is given by
\be
T(M)=\int_k G_{\pi,l}(k)=
T\sum_n\int_\k\frac{1}{E_k^2+\omega_n^2}=T_\tn{div}(M)+T_F(M),
\label{Eq:Tad_pi}
\ee
where $\omega_n=2\pi n T$ and $E_k^2=\k^2+M^2.$ 
Depending on which one of the approximated pion propagators discussed 
in Sec.~\ref{ss:approx} is used, the $T$-dependent mass $M$ 
satisfies either one of the gap equations (\ref{Eq:gap_pole}), 
(\ref{Eq:gap_p0}), or (\ref{Eq:M2_LO}),  or the relation $M^2=h/v.$
Using a 4D cutoff $\Lambda$, the quadratic and logarithmic divergences,
\be
T_d^{(2)}=\frac{1}{16\pi^2}\left[
\Lambda^2-M_0^2\ln\frac{\Lambda^2}{M_0^2} \right],
\quad
T_d^{(0)}=-\frac{1}{16\pi^2}\ln\frac{\Lambda^2}{M_0^2},
\label{Eq:Td2_Td0_def}
\ee
are separated in the divergent part
\be
T_\tn{div}(M)=T_d^{(2)}+[M^2-M_0^2] T_d^{(0)},
\label{Eq:Tad_pi_div}
\ee
by expanding, as in Appendix~A of \cite{patkos08}, the propagator 
$G_{\pi,l}(k)$ around the auxiliary propagator
\be
G_0(p)=\frac{i}{p^2-M_0^2}.
\label{Eq:G0_aux}
\ee
Here, $M_0^2=M_{0B}^2/e$, where $M_{0B}$ is the renormalization scale 
introduced in Sec.~\ref{ss:param}.

The finite part of the tadpole is written as a sum of two terms,
having zero and one statistical factor, respectively
\bea
\nonumber
&&T_F(M)=T_F^0(M)+T^\beta(M),\\
&&T_F^0(M)=\frac{M^2}{16\pi^2}\ln\frac{e M^2}{M^2_{0B}},\quad 
T^\beta(M)=\int\frac{d^3 \k}{(2\pi)^3}\frac{f(E_k)}{E_k},
\label{Eq:Tad_pi_F_decomp}
\eea
where $f(E)=1/(\exp(\beta E)-1).$

The fermion tadpole integral defined in Eq.~(\ref{Eq:T_q_def}) 
is written as
\be 
\tilde T(m_q)=
\frac{T}{N_c}\sum_{i=1}^{N_c}\sum_n\int_\k\frac{1}{E_k^2+\omega_n^2}=
\tilde T_\tn{div}(m_q)+\tilde T_F(m_q),
\label{Eq:T_q}
\ee
where $E_k^2=\k^2+m_q^2,$ $\omega_n=(2 n+1)\pi T-i \mu_i.$ The sum over 
the color degrees of freedom was done with the help of (\ref{Eq:F_P+_P-def}).
The divergent part of $\tilde T(m_q)$ reads
\be
\tilde T_\tn{div}(m_q)=T_d^{(2)}+[m_q^2-M_0^2] T_d^{(0)},
\label{Eq:Tad_q_div}
\ee
while the finite part is decomposed as in (\ref{Eq:Tad_pi_F_decomp}):
\bea
\nonumber
&&\tilde T_F(m_q)=\tilde T_F^0(m_q)+\tilde T^\beta(m_q),\\
&&\tilde T_F^0(m_q)=\frac{m_q^2}{16\pi^2}\ln\frac{e m_q^2}{M^2_{0B}},\quad 
\tilde T^\beta(m_q)=
-\int \frac{d^3 \k}{(2\pi)^3}\frac{1}{2E_k}
\left(\tilde f_\Phi^+(E_k)+\tilde f_{\Phi}^-(E_k)\right),
\label{Eq:Tad_q_F_decomp}
\eea
with $\tilde f_\Phi^\pm(E_k)$ given in (\ref{Eq:F_P+_P-}).

The fermion bubble integral given in Eq.~(\ref{Eq:I_q_def}) is 
decomposed as
\be
\tilde I(p;m_q)=-\frac{T}{N_c}\sum_{i=1}^{N_c}
\sum_n\int_\k\frac{1}{\big[E_1^2+\omega_n^2\big]
\big[E_2^2+(\omega_n-i p_0)^2\big]}=
\tilde I_\textnormal{div}(p;m_q)+\tilde I_F(p;m_q),
\label{Eq:I_q}
\ee
where $E_1^2=\k^2+m_q^2,$ $E_2^2=(\k-\p)^2+m_q^2,$ and 
$\omega_n=(2 n+1)\pi T-i \mu_i.$ 
The divergent part obtained after doing the Matsubara sum 
is independent of the momentum and is given by
\be
\tilde I_\textnormal{div}(p;m_q)=T_d^{(0)},
\label{Eq:I_q_div}
\ee
while the finite part is written again as a sum of two terms,
having zero and one statistical factors, respectively:
\be
\tilde I_F(p;m_q)=\tilde I^0_F(p;m_q)+\tilde I^\beta(p;m_q).
\label{Eq:I_q_finite}
\ee
The finite part with no statistical factor is given by
\bea
\tilde I_F^0(p;m_q)=\frac{1}{16\pi^2}\ln\frac{m_q^2}{M_0^2}+
\frac{Q}{16\pi^2}\left\{
\begin{array}{ll}
\displaystyle
2\arctan Q^{-1},& 0<p^2<4m_q^2, \\
\displaystyle
-\ln\frac{Q-1}{Q+1},& p^2<0, p^2>4 m_q^2, 
\end{array}
\right.
\label{Eq:I_q_0_finite}
\eea
where $Q=\sqrt{\big|1-4m_q^2/p^2\big|},$
while, after summing over color indices, the term with the statistical factor reads
\bea
\displaystyle
\tilde I^\beta(p;m_q)&=&-\int_\k\frac{1}{4 E_1 E_2}
\left[
\frac{\tilde f_\Phi^+(E_1)+\tilde f_\Phi^-(E_2)}
{p_0-E_1-E_2+i\epsilon}-\frac{\tilde f_\Phi^-(E_1)+\tilde f_\Phi^+(E_2)}
{p_0+E_1+E_2+i\epsilon}-\frac{\tilde f_\Phi^+(E_1)-\tilde f_\Phi^+(E_2)}
{p_0-E_1+E_2+i\epsilon}+\frac{\tilde f_\Phi^-(E_1)-\tilde f_\Phi^-(E_2)}
{p_0+E_1-E_2+i\epsilon}
\right].\qquad\ 
\label{Eq:I_q_beta_finite}
\eea

\subsection{The setting-sun integral}
Using the definitions of the fermion bubble integral given in 
(\ref{Eq:I_q_def}) together with the relations (\ref{Eq:I_q}) and 
(\ref{Eq:I_q_div}), the integral defined in (\ref{Eq:J_def}) reads 
in terms of the propagator $D_0$ introduced in (\ref{Eq:D0_prop}) as
\bea
\nonumber
J(M_,m_q)&=&-\frac{1}{N_c}\sum_{i=1}^{N_c}\int_p G^2_{\pi,l}(p) p^2 
\int_q D_0(q) D_0(q+p) + i T_d^{(0)} \int_p p^2 G^2_{\pi,l}(p)\\
&=&\left[1+M^2\frac{d}{d M^2}\right] 
\left(S(M,m_q) -T(M) T_d^{(0)}\right).
\label{Eq:J_decomp}
\eea
Here, we have used that $-i\int_p G_{\pi,l}^2(p)=d T(M)/(d M^2),$ with $M$
satisfying the relation $M^2=h/v$ or the gap-equation (\ref{Eq:M2_LO}) and 
we introduced the setting-sun integral
\be
S(M,m_q)=\frac{1}{N_c} \sum_{i=1}^{N_c} \left[
-i\int_k\int_q D_0(k) G_{\pi,l}(q) D_0(k+q) \right].
\label{Eq:SS}
\ee

With the method described in Refs.~\cite{reinosa04,reinosa06} 
the setting-sun integral can be decomposed as a sum of terms containing zero, 
one, and two statistical factors:
\be
S(M,m_q)=S^0(M,m_q)+S^{\beta,1}(M,m_q)+S^{\beta,2}(M,m_q),
\ee
with
\bse
\bea
S^0(M,m_q)&=&-i\int_k^{(0)}\int_q^{(0)} 
D_0(k) G_{\pi,l}(q) D_0(k+q),
\label{Eq:S0_def}\\
S^{\beta,1}(M,m_q)&=&\frac{1}{N_c} \sum_{i=1}^{N_c}
\int_k^{(0)}\left[\tilde I^0(k;m_q)\sigma(k)
-2 I^0_{q\pi}(k) \tilde\sigma_i(k)\right]
=\left[T^\beta(M)+2 \tilde T^\beta(m_q)\right]T_d^{(0)}+S_F^{\beta,1}(M,m_q),\ \ 
\label{Eq:S1_def}\\
S^{\beta,2}(M,m_q)&=&-\frac{i}{N_c}\sum_{i=1}^{N_c}
\int_k^{(0)} \int_q^{(0)}
\left[\tilde\sigma_i(k)\tilde\sigma_i(q) G_{\pi,l}(k+q)
-2 \tilde\sigma_i(k)\sigma(q) D_0(k+q)
\right],
\label{Eq:S2_def}
\eea
\ese
where $\tilde\sigma_i(k)=\epsilon(k_0)\tilde\rho_0(k)
\big(\tilde f_i^+(|k_0|)+\tilde f_i^-(|k_0|)\big)/2$ and 
$\sigma(k)=\epsilon(k_0)\tilde\rho_0(k) f(|k_0|),$ with
$\epsilon(k_0)$ the sign function, and 
$\tilde\rho_0(k)=2\pi\epsilon(k_0)\delta(k^2-m_q^2)$ and
$\rho_0(k)=2\pi\epsilon(k_0)\delta(k^2-M^2)$ the free spectral 
functions. Here, we have introduced the notation 
$\int_k^{(0)}=\int \frac{d^4 k}{(2\pi)^4}$ and we 
have separated in $S^{\beta,1}$ the divergence coming from (\ref{Eq:I_q}) and
\bea
I^0_{q\pi}(k)=-i\int_q^{(0)}G_{\pi,l}(q) D_0(k+q)=T_d^{(0)}+I^0_{q\pi,F}(k),
\eea
where the finite part is 
\bea
\nonumber
I^0_{q\pi,F}(k)&=&\frac{1}{32\pi^2}\left[
\left(1+\frac{\Delta m^2}{k^2}\right)\ln\frac{M^2}{M_0^2}
+\left(1-\frac{\Delta m^2}{k^2}\right)\ln\frac{m_q^2}{M_0^2}
\right]\\
&&+\frac{G}{16\pi^2 k^2}
\left\{
\begin{array}{ll}
\displaystyle
-\frac{1}{2}\ln\frac{M^2+m_q^2-k^2+G}{M^2+m_q^2-k^2-G} 
-i \pi \Theta(k^2-(M+m_q)^2),&
k^2>(M+m_q)^2, k^2<(M-m_q)^2,\\
\displaystyle
\arctan\frac{k^2+\Delta m^2}{G}+\arctan\frac{k^2-\Delta m^2}{G},&
(M-m_q)^2<k^2<(M+m_q)^2,
\end{array}\right.
\eea
with $\Delta m^2=M^2-m_q^2$ and $G=\sqrt{p^4-2p^2(M^2+m_q^2)+(M^2-m_q^2)^2}.$

In $S^0(M,m_q)$ one expands, as in Ref.~\cite{patkos08}, the propagators 
$G_{\pi,l}$ and $D_0$ around the auxiliary propagator 
defined in (\ref{Eq:G0_aux}) and one obtains
\be
S^0(M,m_q)=S^0(M_0)+
\left[ T^0(M) + 2 \tilde T^0(m_q) -3 T_d^{(2)} \right] T_d^{(0)}
+\left[M^2-M_0^2+2 (m_q^2-M_0^2)\right] T_d^{(I)}+S_F^0(M,m_q),
\label{Eq:S_vac_decomp}
\ee
where 
\be
T_d^{(I)}=-i\int_p^{(0)} G_0^2(p)\left[
-i\int_q^{(0)} G_0(q) G_0(q+p)-T_d^{{0}}
\right]=\frac{1}{3}\frac{d S^0(M_0)}{d M_0^2}-\big[T_d^{(0)}\big]^2-
\frac{T_d^{(0)}}{16\pi^2}.
\label{Eq:TdI}
\ee
The divergent parts of $S^0$ and $S^{\beta,1}$ combine,
so that the complete divergence of the setting-sun integral reads
\be
S_\tn{div}(M,m_q)=S_\tn{div}^{0}(M,m_q)+
S_\tn{div}^{\beta,1}(M,m_q)=S^0(M_0)+
\left[ T(M) + 2 \tilde T(m_q) -3 T_d^{(2)} \right] T_d^{(0)}
+\left[M^2+2 m_q^2-3 M_0^2)\right] T_d^{(I)}.
\label{Eq:S_div}
\ee
Using this expression in (\ref{Eq:J_decomp}) one obtains 
\be
J_\tn{div}(M,m_q)=S^0(M_0)
+\left[2 \tilde T(m_q)-3 T_d^{(2)}\right] T_d^{(0)} +
[2 (M^2+m_q^2)-3 M_0^2] T_d^{(I)}.
\label{Eq:J_div}
\ee

In what follows we give the finite part of the setting-sun integral in
terms of which $J_F(M,m_q)$ can be easily obtained.  With the help of
(\ref{Eq:TdI}) one obtains from (\ref{Eq:S_vac_decomp})
\be
S_F^0(M,m_q)=S^0(M,m_q)-S^0(M_0)-\frac{1}{3}
\left[M^2+2 m_q^2-3 M_0^2\right]\frac{d S^0(M_0)}{d M_0^2}
-\frac{T_d^{(0)}}{16\pi^2}\left[
M^2\ln\frac{M^2}{e M_0^2}+2 m_q^2\ln\frac{m_q^2}{e M_0^2}+3 M_0^2
\right].
\ee
$S_F^0(M,m_q)$ can be most easily calculated numerically as in \cite{yang03} 
by going to Euclidean space and using 4D cutoff $\Lambda$ to regularize 
the integrals. By doing the angular integration one obtains
\be
S^0(M,m_q)=-\frac{1}{2^9\pi^4}\int_0^{\Lambda^2} d x \int_0^{\Lambda^2} d y
\frac{x+y+m_q^2-\sqrt{(x+y+m_q^2)^2-4 x y}}{(x+m_q^2)(y+M^2)},
\ee
with a similar integral for $S^0(M)$, but with $m_q$ replaced by
$M_0.$ Then, by calculating $T_d^{(0)}$ defined in
(\ref{Eq:Td2_Td0_def}) with the same cutoff $\Lambda,$ one can look
for the range of $\Lambda$ where $S_F^0(M,m_q)$ is insensitive to the
variation of $\Lambda.$ In this work we have used 
$\Lambda\in [600,800]$~GeV.

The finite part of the setting-sun integral with one statistical factor 
factorizes upon integration over $k_0$ due to the Dirac delta's of the 
free spectral functions and one has
\be
S_F^{\beta,1}(M,m_q)=\tilde I^0_F(k^2=M^2,m_q) T^\beta(M)+
2 I^0_{q\pi,F}(k^2=m_q^2) \tilde T^\beta(m_q).
\ee 
After performing the frequency and angular integrals in the part of the setting-sun integral containing two statistical factors one obtains
\bea
\nonumber
S^{\beta,2}(M,m_q)&=&-\frac{1}{64\pi^4}\frac{1}{N_c}\sum_{i=1}^{N_c}
\int_0^\infty d |\k|\int_0^\infty |\q| |\k|\,|\q|\bigg\{
\\\nonumber
&&-\frac{2}{E_1 E} \left(\tilde f^+_i(E_1)+\tilde f^-_i(E_1)\right) f(E)
\left[\ln\left|\frac{(E_1+E)^2-E_+^2}{(E_1+E)^2-E_-^2}\right|
+\ln\left|\frac{(E_1-E)^2-E_+^2}{(E_1-E)^2-E_-^2}\right|\ 
\right]\\\nonumber
&&
+\frac{1}{E_1 E_2}\left[\left(
\tilde f_i^+(E_1)\tilde f_i^-(E_2)+\tilde f_i^-(E_1)\tilde f_i^+(E_2)
\right)\ln\left|
\frac{(E_1+E_2)^2-\bar E_+^2}{(E_1+E_2)^2-\bar E_-^2}
\right|\right.
\\
&&\qquad\qquad\left.\left.
+\left(
\tilde f_i^+(E_1)\tilde f_i^+(E_2)+\tilde f_i^-(E_1)\tilde f_i^-(E_2)
\right)\ln\left|
\frac{(E_1-E_2)^2-\bar E_+^2}{(E_1-E_2)^2-\bar E_-^2}
\right|\ 
\right]\right\},
\eea
where $E_1=\sqrt{\k^2+m_q^2},$ $E_2=\sqrt{\q^2+m_q^2},$ $E=\sqrt{\q^2+M^2},$ 
and $E_\pm=\sqrt{(|\k|\pm |\q|)^2+M^2},$ 
$\bar E_\pm=\sqrt{(|\k|\pm |\q|)^2+m_q^2}.$
For the terms containing one fermionic statistical factor the sum over color 
indices will give the functions $\tilde f_\Phi^\pm$ introduced in
(\ref{Eq:F_P+_P-}). For the terms containing two fermionic statistical factors
one introduces the notation $X_\pm=\exp[\beta(E_1\mp \mu_q)]$ and 
$Y_\pm=\exp[\beta(E_2\mp \mu_q)]$ and writes the statistical factors in 
terms of the Polyakov-loop operator.  One obtains
\bea
\nonumber
&&\frac{1}{N_c}\sum_{i=1}^{N_c}\tilde f_i^+(E_1)\tilde f_i^-(E_2)
=\frac{1}{N_c}\sum_{i=1}^{N_c} \tr_c\left[
\frac{1}{L X_+ +1}\frac{1}{L^\dagger Y_- +1}\right]\\
&&=
\frac{1+X_+ Y_-(1+X_+Y_-)+\bar\Phi\big[2Y_-(1+X_+Y_-)+X_+^2\big]
+\Phi\big[2X_+(1+X_+)+Y_-^2\big]+3\Phi\bar\Phi X_+ Y_-}
{(X_+^3+3\bar\Phi X_+^2+3\Phi X_+ +1)(Y_-^3+3\Phi Y_-^2+3\bar\Phi Y_-+1)}.
\label{Eq:f+f-}
\eea
For the term with $\tilde f_i^-(E_1)\tilde f_i^+(E_2)$ one obtains the same 
expression, but with $X_+$ replaced by $Y_+$ and $Y_-$ replaced by $X_-.$ 
Similarly, one has
\bea
\nonumber
&&\frac{1}{N_c}\sum_{i=1}^{N_c}\tilde f_i^+(E_1)\tilde f_i^+(E_2)\\
&&=\frac{3\bar\Phi^2 X_+^2 Y_+^2+6\Phi^2 X_+Y_++(3\Phi\bar\Phi-1)X_+Y_+(X_++Y_+)
+2\Phi(X_++Y_+-X_+^2 Y_+^2)+\bar\Phi(X_+-Y_+)^2+1}
{(X_+^3+3\bar\Phi X_+^2+3\Phi X_+ +1)(Y_+^3+3\bar\Phi Y_+^2+3\Phi Y_+ +1)},
\label{Eq:f+f+}
\eea
and for the term with $\tilde f_i^-(E_1)\tilde f_i^-(E_2)$ one has the
same expression, but with $\bar\Phi$ interchanged with $\Phi,$ and
$X_+$ replaced by $X_-$ and $Y_+$ replaced by $Y_-.$ We note here
that, as one can see by setting $\Phi=\bar\Phi=0,$ the expression on
the right-hand side of (\ref{Eq:f+f+}) can be negative and in consequence 
it does not allow for an interpretation in terms of distribution functions.

When the approximation $G(k)\to D(k)$ is used, the setting-sun
integral defined in (\ref{Eq:SS}) appears also in the expression of
the quark-pion two-loop integral of (\ref{Eq:Omega_grand_pot}) which 
reads
\bea
I_2=-\sqrt{N}\frac{g^2}{2}i\tr_{D,c}\int_k\int_p
\gamma_5 D(k)\gamma_5 D(k+p)G_{\pi,l}(p)
=g^2\sqrt{N} N_c \left[\tilde T_2(m_q)-2\tilde T(m_q) T(M) - M^2 S(M,m_q)
\right],\ \ 
\label{Eq:I2}
\eea
where depending on the approximation on the pion propagator  $M$ 
satisfies either one of the gap equations (\ref{Eq:gap_pole}), 
(\ref{Eq:gap_p0}), (\ref{Eq:M2_LO})  or the relation $M^2=h/v.$ 
Here, we introduced the integral
\be
\tilde T_2(m_q)=\frac{1}{N_c}\sum_{i=1}^{N_c} \left[
-i\int_k\int_q D_0(k) D_0(q) \right],
\label{Eq:T2_def}
\ee
which, after performing the Matsubara sums, can be written as
\bse
\bea
&&\tilde T_2(m_q)=\tilde T_2^0(m_q)+\tilde T_2^{\beta,1}(m_q)+
\tilde T_2^{\beta,2}(m_q),\\
&&\tilde T_2^0(m_q)=\big(\tilde T(m_q)\big)^2,\qquad
\tilde T_2^{\beta,1}(m_q)=2\tilde T^0(m_q) \tilde T^\beta(m_q)\\
&&\tilde T_2^{\beta,2}(m_q)=
\frac{1}{16\pi^4}
\int_0^\infty d |\k| \int_0^\infty d |\q| \frac{k^2 q^2}{E_1 E_2}
\sum_{r,s=\pm}\left[
\frac{1}{N_c}\sum_{i=1}^{N_c}\tilde f_i^r(E_1)\tilde f_i^s(E_2)\right],
\eea
\label{Eq:T2_decompose}
\ese
where $E_1=\sqrt{\k^2+m_q^2},$ $E_2=\sqrt{\q^2+m_q^2},$ and the color sums 
are given in (\ref{Eq:f+f-}) and (\ref{Eq:f+f+}). In terms of the quantities 
defined in (\ref{Eq:T2_decompose}) the derivative with respect to $\Phi$ 
of $I_2$ appears in (\ref{Eq:dU_dPhi}).

}
\end{document}